\documentclass[aps,pra,twocolumn,superscriptaddress,nofootinbib,10pt]{revtex4-2}

\usepackage[T1]{fontenc}
\usepackage{amsmath}      
\usepackage{amssymb}
\usepackage{graphicx}
\usepackage{mathrsfs}
\usepackage{multirow}
\usepackage{comment}
\usepackage{soul}         
\usepackage{braket}
\usepackage{xcolor}
\usepackage{tensind}
\usepackage{pifont}       

\usepackage{orcidlink}

\hypersetup{
	colorlinks=true,
	allcolors=blue
}

\newcommand{\eg}{\textit{e.g.}}

\def\svapo{\texttt{GINTONIC}}
\def\mbf{\boldsymbol}

\def\Om{\boldsymbol{\Omega}}

\definecolor{cyan}{rgb}{0,0.9,0.9}
\definecolor{orange}{rgb}{0.9,0.5,0}
\definecolor{magenta}{rgb}{1,0,1}
\definecolor{purple}{rgb}{0.8,0.4,0.8}
\definecolor{gray}{rgb}{0.8242,0.8242,0.8242}

\begin{document}

\title{Dynamics of quantized vortices under quasi-periodic boundary conditions}

\author{Fabio Magistrelli\,\orcidlink{0009-0005-0976-7851}}
\email{fabio.magistrelli@uni-jena.de}
\affiliation{Theoretisch-Physikalisches Institut, Friedrich-Schiller-Universit{\"a}t Jena, 07743 Jena, Germany}

\author{Marco Antonelli\,\orcidlink{0000-0002-5470-4308}}
\email{antonelli@lpccaen.in2p3.fr}
\affiliation{CNRS/IN2P3, Laboratoire de Physique Corpusculaire de Caen (LPC), 14050 Caen, France}

\date{\today}

\begin{abstract}
The Gross--Pitaevskii equation is widely used for vortex dynamics, but finite domains with hard walls or confining potentials distort bulk behavior through vortex-image effects or induced flows. Periodic boundaries reduce wall artifacts yet cannot realize finite net vorticity because of topological obstruction, so bulk simulations with non-zero circulation are typically unavailable. Hence, we impose quasi-periodic boundary conditions that keep the superfluid's density periodic while enforcing phase windings consistent with a net prescribed total vorticity. This setting conserves the net number of vortices and enables long-time tracking of vortex trajectories in settings that finite containers cannot capture. This allows us to study vortex depinning and nucleation leading to the creation of K\'arm\'an vortex streets and perfectly periodic vortex arrays. The framework also provides a toy model for studying vortex dynamics in the bulk of neutron stars, free of possible limitations induced by confining potentials.
\end{abstract}

\maketitle


\section{Introduction}
\label{sec:intro}

A common strategy to mimic the bulk behavior of an extended system is to impose periodic boundary conditions (PBC).
This approach is widely used, from molecular dynamics to studies of quantum turbulence \cite[\eg][]{Kondaurova2008,Krstulovic2011PhRvE,Baggaley2018PhRvA}, and even to dynamical processes in the inner crust of neutron stars~\citep{Sekizawa2022PhRvC,almirante2024slab,Almirante2024rod,pecak2024toolkit,Martin:2016ubw}.
Unlike box boundaries, the toroidal topology provides a closed system that can sustain homogeneous flows, making it a natural choice for investigating the thermodynamic properties of superfluids with persistent currents~\citep{andreev2004JLTP,gavassino2020}.

However, standard PBC suffer from a fundamental limitation: they forbid any net vorticity in the computational domain, since vortices cannot consistently pierce a torus without violating the single-valuedness of the order parameter~\citep{kleinert2008multivalued,Turner_RMP}.
This topological obstruction prevents the use of ordinary PBC to study fluid elements containing a finite, net number of vortices, even though such systems are relevant both as idealised bulk limits of laboratory systems and, even more, for fluid elements in neutron stars~\citep{GraberNSlab,amp_review_2023}.

A way forward is provided by the quasi-periodic boundary conditions (QPBC) introduced in two dimensions in~\citet{Mingarelli_2016,Mingarelli2018} and later extended to three dimensions by~\citet{WoodTobyS.2019Qbcf}.
QPBC generalise PBC by allowing the condensate wave function to vary across the domain, such that the density remains periodic while the phase carries a finite circulation around the boundaries.
QPBC also provide an alternative to hard-wall domains currently used in studies of the thermodynamics of fluid elements with net vorticity~\citep{pekac_chamel_2021}, and in simulations of vortex–nucleus~\citep{bulgac2013prl,Warszawski_unpinning,klausner_pinnning_2023} and vortex–fluxtube pinning forces~\citep{thong2023MNRAS}. The advantage of using the QPBC in this case is to remove spurious vortex-image effects or hard-walls induced flows.
The natural application of QPBC is the study of infinite vortex lattice geometries and energetics~\citep{Doran2020}, including stationary vortex lattices in multi-component systems~\citep{WoodGraber}.

Simulating superfluids in a closed domain that supports both finite vorticity and persistent current enables more realistic extensions of vortex-filament simulations in a pinning landscape~\citep{AntoHaskell,link2022ApJ}.
These studies demonstrated the transition from weak mutual friction (when vortices remain pinned) to a linear HVBK-like regime (when unpinned), while also exploring the role of different pinning landscapes, with possible implications for pulsar glitches~\citep{amp_review_2023}.
Periodic boundaries were crucial, as they allowed vortices to be tracked indefinitely without leaving the system, and the random pinning landscape to be generated once in a finite region rather than in real time around moving vortices.
More recently, \citep{Stockdale_PRA_21,Liu2024} extended this framework to dynamical simulations of the Gross--Pitaevskii equation (GPE), investigating vortex depinning under an imposed external current.
Here, motivated by the point-vortex study in \citep{AntoHaskell}, we develop and test a numerical complex-time framework (real-time with phenomenological dissipation) that mimics the bulk behavior of a superfluid, where a fluid element can be immersed in persistent homogeneous currents while also being pierced by quantized vortices.

In this paper, we introduce \svapo\footnote{
	\url{https://github.com/Magistrelli/gintonic}},
a C++ code for solving the two-dimensional GPE with QPBC in the presence of arbitrary, time-dependent external potentials.
Although GPE simulations with PBC are numerous, the QPBC regime remains essentially unexplored, except for the depinning simulations of \citet{Liu2024} and the vortiex-lattice formation simulations of~\citep{Doran2020}. Since our numerical approach differs from the one presented in \citep{Doran2020}, validating the GPE in this QPBC setting requires a series of benchmark tests, which are nevertheless needed to establish reliability.
This, in turn, opens the door to revisiting already studied phenomena (\eg~the K\'arm\'an vortex street~\citep{SasakiKarmanGP}) in the presence of finite net vorticity and no hard walls nor confining potentials.

The paper is organized as follows.
In Sec.~\ref{sec:methods}, we recall the main properties of the GPE. We also motivate and describe the QPBC needed to model systems with non-zero net vorticity.
We present some first applications, including vortex lattice formation and vortex nucleation from moving impurities, in Sec.~\ref{sec:numSim}.
We summarize our results in Sec.~\ref{sec:conclusions}.
In Appendix~\ref{app:numChecks}, we validate the implementation of \texttt{GINTONIC} by reproducing the expected behavior of simple vortex systems.

\section{Fluid elements with non-zero net vorticity}
\label{sec:methods}

This section follows the discussion and ideas presented in \citet{Mingarelli_2016,WoodTobyS.2019Qbcf}, rewriting them in a time-dependent scenario. The aim is to develop a framework that mimics the bulk behavior of a superfluid, namely a fluid element where persistent currents can be homogeneous and stationary, and where quantised vortices can be traced for long times without leaving the finite computational domain~\citep{AntoHaskell,Liu2024}.

\subsection{General framework}
\label{subsec:gpe}

Consider a complex field $\psi$ whose dynamics is governed by the GPE~\citep{gross1961structure, pitaevskii1961vortex},
\begin{equation}
	\label{eq:GPE}
	\partial_t \psi =  \left[ g |\psi|^2 + V(\boldsymbol{x}, t) - \mu - \frac{\hbar^2 \nabla^2}{2 m} \right] \psi \, ,
\end{equation}
where $m$, $g$, and $\mu$ are positive parameters, $V(\boldsymbol{x}, t)$ is a time-dependent external potential, and $\hbar$ is the reduced Planck constant.
When using the GPE to effectively describe Bosonic condensates (or even Fermionic superfluids~\citep{Salasnich2009}), $m$ is interpreted as the mass of the constituent bosons, $g$ as the inter-particle interaction coupling (that we take to be repulsive), and $\mu$ as the chemical potential.
The healing length $\xi = \hbar/\!\sqrt{m \mu}$ sets a physical scale for the system, and a natural scale for  $|\psi|^2$ is the bulk density $n_\infty= \mu / g$.
Combining $\xi$ with the first sound velocity, $c_1 = \sqrt{\mu/m}$, we can define the fundamental timescale~$\tau = \hbar/\mu$~\citep{pethick_book_2008}.

The GPE is sometimes employed as an analogue model for quantized vortices in the inner crust of neutron stars \cite[\eg][]{Warszawski_GPE_2011,Warszawski_unpinning,Liu2024,Liu2025}.
In this context, where vortices are immersed in a non-homogeneous environment, the inclusion of the external potential $V$ provides the pinning landscape that is crucial to simulate the stick–slip dynamics of vortices arising from their interaction with pinning centres~\citep{haskell2016MNRAS,AntoHaskell,link2022ApJ}.
Here the superfluid consists of neutron Cooper pairs and the interaction is mediated by nuclear forces, with typical scales $m \sim 2\, \rm{GeV}$, $\xi \sim 3\,[\mu/\rm{MeV}]^{-1/2}\, \rm{fm}$, and $\tau \sim 6\,[\mu/\rm{MeV}]^{-1} \times 10^{-22}\, \rm{s}$~\citep{GraberNSlab,AntoHaskell}.
For comparison, when the GPE is used as an effective model for vortex dynamics in the bosonic He-II, the relevant scales are $m \sim 4\, \rm{GeV}$, $\xi \sim 0.1\,[\mu/\rm{meV}]^{-1/2}\, \rm{nm}$, and~$\tau \sim 6\,[\mu/\rm{meV}]^{-1} \times 10^{-13}\, \rm{s}$ \cite[\eg][]{Roberts_1971,Barenghi2001bookinto}.
In what follows, we adopt natural units with $\hbar = m = 1$, while $\mu$ sets the energy scale. Times and lengths are expressed in units of $\tau$ and~$\xi$.

For our purposes, we consider the GPE in a rotating frame with angular velocity~$\mbf{\Om}$,
\begin{multline}
	\label{eq:origGPE}
	(i-\gamma)\, \partial_t \psi = \left[ \frac{|\psi|^2}{n_\infty} + \frac{V(\boldsymbol{x}, t)}{\mu} - 1 - \frac{\nabla^2}{2} + \right. 
    \\
     + i \Om \cdot \left(\boldsymbol{x}\times\mbf{\nabla} \right) \Big] \psi \,,
\end{multline}
where $\gamma \geq 0$ regulates dissipation~\citep{Pitaevskii:1958, Choi:4057, Tusbota:2002, KasamatsuTsubota, Doran2020}. Setting $\gamma >0$ breaks unitarity and makes the condensate evaporate.

In the following, Eq.~\eqref{eq:origGPE} will always be defined over finite domains with $2\Om = \langle\mbf{\omega}_0\rangle$, where the mean net vorticity in the domain is defined as
\begin{equation}
	\label{wGlobMean}
	\langle\mbf{\omega}_0\rangle
	= \frac{h}{m} \frac{\boldsymbol{N}_v}{V_0} \,.
\end{equation}
Here, $\boldsymbol{N}_v = \sum_i s_i \boldsymbol{l}_i$ is the oriented number of vortices contained in the domain of volume $V_0$, weighted by their charges $s_i$ and lengths.
The vector $\boldsymbol{l}_i$ connects the two points where the $i$-th vortex line intersects the fundamental cell, and $\boldsymbol{N}_v$ is thus a length. 
For a uniformly rotating condensate, combining $2\Om = \langle\mbf{\omega}_0\rangle$ with Eq.~\eqref{wGlobMean} expresses the Feynman-Onsager relation~\citep{feynman1955chapter}.

Consider now a perfectly periodic vortex-array configuration of $\psi$. 
Due to periodicity, vortex lines piercing the fundamental cell must close on themselves after a finite number of cell crossings. This configuration allows us to simulate finite domains with a non-zero $\langle\mbf{\omega}_0\rangle$, as described in \citet{Mingarelli_2016,WoodTobyS.2019Qbcf}.
The net vorticity must be conserved.
On the one hand, the Kelvin circulation theorem remains valid also in a fundamental cell of the lattice, thus preventing non-contractible vortices from originating inside the domain \citep{ArnoldKhesin} (see also Appendix~\ref{app:vortEq_HelmTheo}). 
On the other hand, the boundary conditions in the fundamental cell ensure that already existing vortices cannot modify $\boldsymbol{N}_v$, as for each vortex (or part of it) leaving the cell, another one must enter from the opposite side.

\subsection{Quasi-periodic boundary conditions}
\label{subsec:qpbc}

The following discussion is largely based on~\citep{tkachenko1966vortex,Mingarelli_2016,Mingarelli2018,WoodTobyS.2019Qbcf,Doran2020}.
In an infinite vortex lattice, the density $n=|\psi|^2$ is periodic, while the phase $\theta = \arg\psi$ is quasi-periodic due to possible branch cuts crossing the boundaries of the fundamental cell.
\citet{tkachenko1966vortex} realised that $\psi$ can be expressed in terms of the first Jacobi theta function,
\begin{equation}
	\label{eq:JacobiTheta}
	\Theta_1 \left(z,q\right)
	\equiv 2 \sum_{n=0}^{\infty} (-1)^n\, q^{\left(n+\frac{1}{2}\right)^2} \sin{\left[(2n+1)z\right]} \,,
\end{equation}
where $z$ is a complex coordinate and $q$ is a complex parameter describing the periodicity of its singularities \citep{JacTheta}. Given a 2D Bravais vortex lattice with fundamental vectors $\boldsymbol{\lambda}_1 = \pi \hat{\mbf{e}}_x$ and $\boldsymbol{\lambda}_2 = \lambda_{2,x} \hat{\mbf{e}}_x + \lambda_{2,y} \hat{\mbf{e}}_y$, one can always define a fundamental cell by taking its sides as $\boldsymbol{L}_i = \boldsymbol{\lambda}_i$, so that $N_v = 1$. In this case, the phase of $\psi$ can be written as~\citep{WoodTobyS.2019Qbcf}
\begin{equation}
	\label{eq:preWoodTheta}
	\begin{split}
		\theta_J(\boldsymbol{x}|\boldsymbol{x}_v) = & \, \frac{x(y-2y_v)}{\lambda_{2,y}} \\
		& + \text{Arg} \left[ \Theta_1 \left(x-x_v + i(y-y_v) \,,\, e^{i\pi\tau} \right) \right] \,,
	\end{split}
\end{equation}
where $\tau \equiv (\lambda_{2,x} + i\,\lambda_{2,y}) / \pi$ and $\boldsymbol{x}_v \equiv (x_v, y_v)$ position of the vortex closest to the rotational axis, always forced to be in $\boldsymbol{x} = 0$ by the first term on the right hand side.
Equation~\eqref{eq:preWoodTheta} can describe any simple Bravais lattice, as any couple of vectors can always be mapped into $(\boldsymbol{\lambda}_1$, $\boldsymbol{\lambda}_2)$ by rescaling the coordinate vector by $\pi|\boldsymbol{\lambda}_1|^{-1}$ and rotating it clockwise by the angle $\phi = \arctan(\lambda_{1,y}/\lambda_{1,x})$.

We can describe any periodic array by considering the Bravais lattice associated with each vortex of the fundamental cell separately, and then summing their single contributions given by Eq.~\eqref{eq:preWoodTheta}, provided that every vortex is weighted in the sum by its charge $s$.
The phase of a periodic array of vortices can thus be written as
\begin{equation}
	\label{eq:WoodTheta}
	\theta(\boldsymbol{x}) = \sum_k s_k\, \theta_J(\boldsymbol{x}|\boldsymbol{x}_k) \,,
\end{equation}
where $k$ runs over all the vortices in the fundamental cell and $\boldsymbol{x}_k$ are their positions.
The phase in Eq.~\eqref{eq:WoodTheta} satisfies the QPBC of \citet{WoodTobyS.2019Qbcf},
\begin{equation}
	\label{eq:WoodPbc}
	\theta(\boldsymbol{x}+\boldsymbol{L})
	= \theta(\boldsymbol{x}) + \frac{\pi}{V_0} (\boldsymbol{N}_v\times\boldsymbol{L})\cdot\boldsymbol{x} + c \mod{2\pi} \,,
\end{equation}
where $\boldsymbol{L}$ represents any discrete translational symmetry of the vortex array. In the static case, $c$ is an arbitrary integration constant, specific for each $\boldsymbol{L}$, that fixes the projection of the mean position of the vortices on the direction of that translational symmetry.
In the case of zero net vorticity, $\boldsymbol{N}_v = 0$, the QPBC reduce to the usual PBC. In this case, since we are in the reference frame of the superfluid, the integration constants must be zero, as otherwise they would introduce a uniform translational velocity.

In a $D$-dimensional case, any selected symmetry vector can always be expressed as a linear combination of the fundamental cell's side vectors $\boldsymbol{L}_i$ as $\boldsymbol{L} = \sum_i l_i \boldsymbol{L}_i$, where $i = 1,\dots,D$ and $l_i \in \mathbb{Z}$.
The QPBC in Eq.~\eqref{eq:WoodPbc} can thus be extended to the dynamic case as
\begin{equation}
	\label{eq:prePbc}
	\begin{split}
		\theta(t)|_{\boldsymbol{x}}^{\boldsymbol{x}+\boldsymbol{L}} =
		& \,\, \frac{\pi}{V_0} (\boldsymbol{N}_v\times\boldsymbol{L})\cdot\boldsymbol{x} + \sum_i l_i c_i(t) \\
		& + \frac{\pi}{V_0} \sum_{i,j>i} \,l_i\,l_j \,\boldsymbol{L}_i\cdot(\boldsymbol{N}_v\times\boldsymbol{L}_j) \mod{2\pi} \,,
	\end{split}
\end{equation}
where $c_i(t)$ are now time-dependent spatial-integration constants specific for each side vector.
Equation~\eqref{eq:prePbc} holds for any shape of the fundamental cell, and it does not require the side vectors to be perpendicular to each other.
By construction, the phase in Eq.~\eqref{eq:WoodTheta}, and thus the QPBC in Eq.~\eqref{eq:prePbc}, are compatible with $\Om = \langle \Om_0 \rangle$ in Eq.~\eqref{eq:origGPE}. 
The mutual compatibility of the QPBC in Eq.~\eqref{eq:prePbc} between the different directions $\boldsymbol{L}_i$ can be proved, at each fixed time $t$, in the same way as \citet{WoodTobyS.2019Qbcf} did for the static case.

At a fixed time, the integration constants $c_i(t)$ of Eq.~\eqref{eq:prePbc} can be expressed in terms of the phase field evaluated \eg~in $\boldsymbol{x} = \pm \boldsymbol{L}_i / 2$ as
\begin{equation}
	\label{eq:ciFromTheta}
	c_i(t) = \theta(\boldsymbol{L}_i/2,t)-\theta(-\boldsymbol{L}_i/2,t) \,.
\end{equation}
The time evolution of the integration constants $c_i$ is therefore determined by the evolution of the phase field, and it is thus prescribed by the GPE.
The quasi-periodicity of Eq.~\eqref{eq:origGPE} assures the derivative of the phase field to be periodic.
As a consequence, the $c_i$ are constant in time and do not need to be explicitly evolved.
However, allowing for the time evolution of the integration constants allows us to check this result as a sanity check.
Moreover, it effectively generalizes the QPBC to also describe a net superfluid flow, which would instead require the inclusion of Bloch boundary conditions.

In our code, we solve the GPE for the real and imaginary parts of the superfluid wave function $\psi = R + iI$. We write the boundary conditions for $R$ and $I$ by combining the QPBC of Eq.~\eqref{eq:prePbc} with simple PBC for the number density $|\psi|^2$,
\begin{equation}
	\label{eq:qpbc}
    \begin{split}
	R(\boldsymbol{x}+\boldsymbol{L}, t) = R(\boldsymbol{x}, t) A(\boldsymbol{x}, t) - I(\boldsymbol{x}, t) B(\boldsymbol{x}, t) \,, \\
	I(\boldsymbol{x}+\boldsymbol{L}, t) = I(\boldsymbol{x}, t) A(\boldsymbol{x}, t) + R(\boldsymbol{x}, t) B(\boldsymbol{x}, t) \,,
    \end{split}
\end{equation}
where $A$ and $B$ are given by
\begin{align}
	A(\boldsymbol{x}, t) = \cos{\! \left[ \theta(t)|_{\boldsymbol{x}}^{\boldsymbol{x}+\boldsymbol{L}} \right]} \,, \\
	B(\boldsymbol{x}, t) = \sin{\! \left[ \theta(t)|_{\boldsymbol{x}}^{\boldsymbol{x}+\boldsymbol{L}} \right]} \,,
\end{align}
with $\theta(t)|_{\boldsymbol{x}}^{\boldsymbol{x}+\boldsymbol{L}}$ given by Eq.~\eqref{eq:prePbc}.

\subsection{Ansatz for the initial condition}\label{subsec:initCond}

In a vortex-free case, neglecting $\nabla^2 \psi $ in the GPE gives 
\begin{equation}
	\label{eq:depletion_pot}
	n(\boldsymbol{x}) = n_\infty f_\textrm{bg}(\boldsymbol{x}) \,,
\end{equation}
where $n = |\psi|^2$ and $f_\textrm{bg}(\boldsymbol{x}) = 1 - V(\boldsymbol{x})/\mu$. This can be used as a reasonable ansatz for the initial condition in the absence of vortices. Note that it is also possible to shift $V$ by a constant (effectively changing $\mu$) to satisfy
\begin{equation}
	\label{eq:normV}
	\int V(\boldsymbol{x}) \, d^2x = 0 \,,
\end{equation}
which ensures that the average density in the fundamental cell is~$\langle n \rangle = n_\infty$. 

In the infinite plane, the depletion caused by a single vortex of charge $s$ can be approximated as $n(r) = n_\infty f_s^2(r)$, with $f_s(r) = r/(2s^2+r^2)^{1/2}$ and $r$ the radial distance from the vortex axis \citep{pethick_book_2008}. 
The total depletion caused by multiple vortices can then be approximated as
\begin{equation}
	\label{eq:depletion_vort}
	n(\boldsymbol{x}) = n_\infty \, \prod_{k} f_s^2(|\boldsymbol{x} - \boldsymbol{x}_k|) \,,
\end{equation}
where $k$ runs over all vortices in the array and $\boldsymbol{x}_k$ are their positions.

At every time during the system's dynamical evolution, the density $|\psi|^2$ and phase $\arg \psi$ fields must satisfy, respectively, the PBC and QPBC introduced in Eq.~\eqref{eq:prePbc}. 
To imprint vortices at specified positions at $t=0$, we initialize the phase field with Eq.~\eqref{eq:WoodTheta}, truncating the sum in Eq.~\eqref{eq:JacobiTheta} at $n=30$. 
The initial superfluid density is obtained by combining Eq.~\eqref{eq:depletion_pot} and Eq.~\eqref{eq:depletion_vort}, so that the initial condition reads
\begin{multline}
	\label{eq:init_cond}
	\psi(\boldsymbol{x},t=0)
	= \sqrt{n_\infty f_\text{bg}(\boldsymbol{x})} \prod_{k=1}^{N} f_s(|\boldsymbol{x}-\boldsymbol{x}_k|) 
	\\
	\exp\{i s_k \theta_{\!J}(\boldsymbol{x}|\boldsymbol{x}_k)\} \,,
\end{multline}
where $\boldsymbol{x}_k$ and $s_k$ are the positions and charges of the vortices in the fundamental cell. 
The corresponding density is
\begin{equation}
	\label{eq:initial_dens}
	n(\boldsymbol{x}, t=0) = n_\infty \, f_\textrm{bg}(\boldsymbol{x}) \, \prod_{k=1}^N f_s^2(|\boldsymbol{x} - \boldsymbol{x}_k|) \,.
\end{equation}
The ansatz relies on the fact that the modulation of the density $n$ caused by $V$ and the total depletion in the core of a vortex are essentially independent.
It produces reasonable initial configurations of $\psi$ that, however, contain fast propagating modes injected in the system at $t=0$. 
A more polished initial configuration can be achieved by evolving the system with $\gamma > 0$, and then switching to $\gamma = 0$ once the spurious sound has dissipated.
The integration constants $c_i(t=0)$ are initialized using Eq.~\eqref{eq:ciFromTheta}.

\section{Numerical simulations}
\label{sec:numSim}

We now study a selection of physical scenarios that can be simulated within our QPBC framework.
In \svapo, we solve Eq.~\eqref{eq:origGPE} on a 2D rectangular Cartesian grid with the new time-dependent QPBC introduced in Eq.~\eqref{eq:prePbc}. The field $\psi$ is decomposed into real and imaginary parts and evolved in time with a third-order Adams–Bashforth linear multistep method~\citep{Butcher:2016}. Spatial derivatives are computed with five-point stencils.\footnote{For comparison and validation, the code also includes a simpler Euler solver and a fourth-order explicit Runge–Kutta method, together with three-point stencils for spatial derivatives.}

All results shown here use the natural units introduced in Sec.~\ref{subsec:gpe}. 
Unless otherwise specified, we set $n_\infty = \mu = 1$, use a time resolution of $\delta t = 5 \times 10^{-4}\,\tau$ (small enough to always satisfy the CFL condition for the GPE), and perform all simulations in the non-dissipative regime,~$\gamma = 0$.

\subsection{Vortex lattice formation under QPBC}

Effective dissipation of vortex motion in GPE simulations (whether due to discretization \citep{Gecse2005prb}, physical, or phenomenologically modeled via $\gamma>0$) leads to the formation of a vortex lattice. 
Several numerical experiments demonstrated this effect in superfluids confined by rotating traps \cite[\eg][]{Tusbota:2002,Lobo_PRL_2004}. However, the geometry imposed by the trap introduces finite-size effects, which can produce frustrated configurations if the computational domain is not large enough~\cite[\eg][]{Adhikari2019}.

In principle, exact lattice configurations with $N_v \neq 0$ are accessible only with the QPBC of Eq.~\eqref{eq:prePbc}, see~\citep{Mingarelli_2016,Mingarelli2018,WoodTobyS.2019Qbcf,Doran2020} and references therein. 
Following \citep{Doran2020}, we therefore investigate the dissipative process of lattice formation starting from an array of vortices randomly distributed in the fundamental cell. 
We run a series of simulations with a finite dissipative parameter, $\gamma=0.03$, $\delta l \simeq 0.3$, and exploring different cell geometries, background potentials, and vortex densities $N_v$.
To remain consistent with the use of QPBC, we consider static potentials of the form
\begin{equation}
	\label{eq:GaussPeak}
	V(\boldsymbol{x}) = \sum_{n,m \in \mathbb{Z}} V_0 \exp\left( -\frac{|\boldsymbol{x} - \boldsymbol{x}_{n,m}|^2}{2 \sigma^2} \right) \,,
\end{equation}
where $\boldsymbol{x}_{n,m} = \boldsymbol{x}_{0,0} + n L_x \hat{\boldsymbol{x}} + m L_y \hat{\mbf{y}}$, with $L_x$ and $L_y$ fundamental periodicity vectors of the vortex array (i.e.~the dimensions of the numerical cell where the QPBC are imposed).
The strength of the potential is set by the real constant $V_0$, and its width by the positive parameter $\sigma$.

We also perform runs with $\gamma=0$ but with resolution $\delta l = \xi$ low enough to introduce numerical dissipation. 
Our code remains stable at this and lower resolutions, and qualitatively reproduces the same results as the dissipative runs. 
For all simulations in this section, the chemical potential is fixed to $\mu=0.5$. 
The time resolution is set by the CFL condition with Courant number $0.1$ in the dissipative runs, and fixed to $5\times10^{-4}\,\tau$ in the low-resolution ones.

\begin{figure*}[t]
	\centering 
	\includegraphics[width=\textwidth]{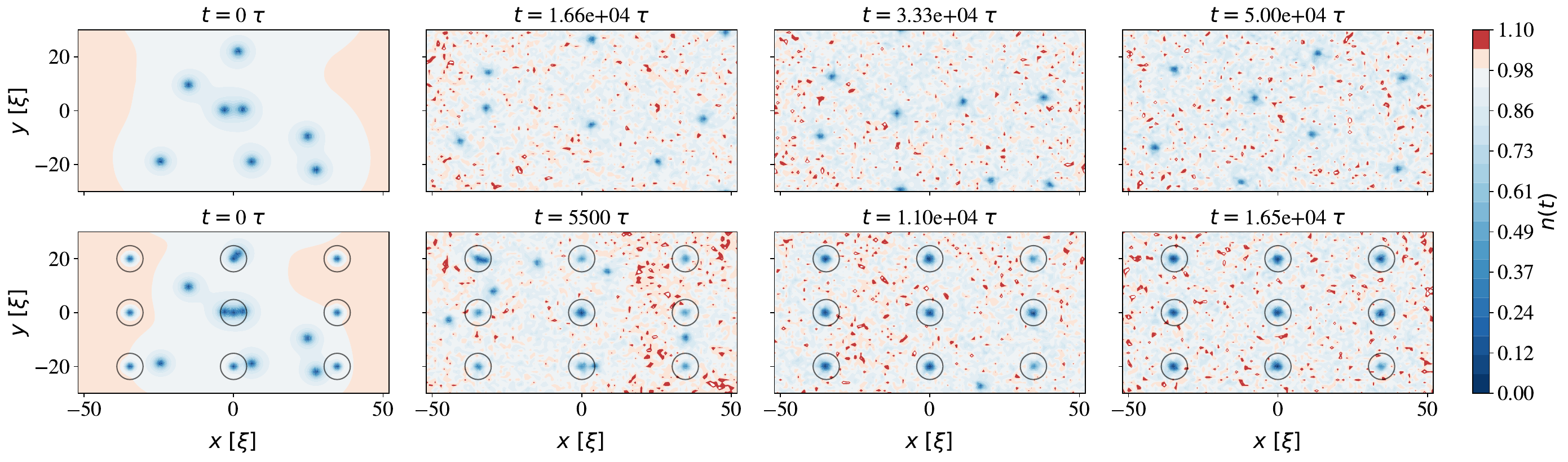}
	\caption{Formation of vortex lattices in a rectangular cell due to numerical dissipation.
		The panels show a series of superfluid-density snapshots without (\textit{top row}) and with (\textit{bottom row}) a background potential given by a lattice of Gaussian peaks. The initial positions of the $N_v = 8$ vortices are clearly visible in the first panel of the top row as points of complete depletion. They are identical in the two simulations. In the bottom row, black circles mark the (fixed) peaks of the background potential. In the homogeneous ($V=0$) system, the vortices form a triangular lattice by the end of the run. With the external potential, they are forced to pin on the peaks, producing a frustrated rectangular lattice. The empty peak at the bottom right of the fundamental cell remains visible because the superfluid depletion there is incomplete.}
	\label{fig:lattice}
	\includegraphics[width=\textwidth]{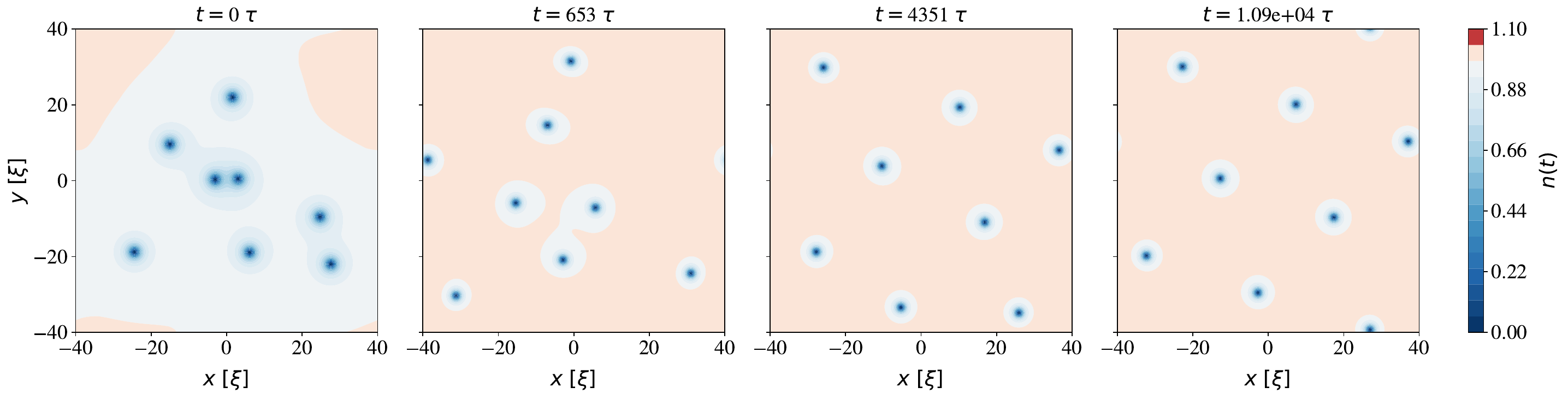}
	\caption{Formation of a vortex lattice in a square cell using the dissipative GPE with $\gamma = 0.03$. At late times, the $N_v=8$ vortices form an Abrikosov lattice.}
	\label{fig:lattice_square}
\end{figure*}

In Fig.~\ref{fig:lattice}, we examine lattice formation for a system with $N_v=8$ in a rectangular cell of size $60\sqrt{3}\,\xi \times 60\,\xi$ with numerical dissipation. 
The cell geometry is chosen to be compatible with an Abrikosov (triangular) lattice of vortices, i.e.~the energetically most favorable configuration. 
We consider both an homogeneous ($V=0$) superfluid and one immersed in the Gaussian background potential of Eq.~\eqref{eq:GaussPeak}, with $V_0 = \mu$, $\sigma = \xi$, and periodicity $3n,3m \in \mathbb{Z}$. 
The initial vortex positions are random within the computational cell.

In the homogeneous case ($V=0$), the system relaxes toward an Abrikosov lattice. At late times, it explores a family of metastable triangular configurations that differ only by continuous rigid translations. In contrast, when the external potential is present, dissipation drives vortex pinning on the Gaussian peaks. The long-time state is then a frustrated configuration, where the vortices occupy the sites of the rectangular lattice imposed by the potential, leaving one Gaussian peak unfilled.

Figure~\ref{fig:lattice_square} shows the evolution of an analogous system with $N_v=8$ vortices in a square cell of side $L = 80\,\xi$, for an homogeneous ($V=0$) superfluid now evolved with analytical dissipation $\gamma = 0.03$. The system slowly relaxes to its ground state, again an Abrikosov lattice.

\subsection{K\'arm\'an vortex street generated by an initially pinned vortex~($N_v=1$)}
\label{subsec:depKarm}

The use of QPBC makes it possible to study depinning and nucleation of vortices in a periodic background potential, such as the one in Eq.~\eqref{eq:GaussPeak}. 
A numerical experiment in this direction was recently proposed in \citet{Liu2024}, where the depinning of a vortex from a Gaussian potential under an increasing external current was analyzed (both with QPBC, as well as with other boundary conditions).

Here, we follow the same idea and consider systems with $N_v=1$, where the QPBC are particularly advantageous and allow us to implement the GPE extension of the point-vortex simulations presented in~\citep{AntoHaskell}. While the latter can only be used to study how the depinning threshold depends on the properties of the pinning center, the GPE extension also allows us to investigate the possibility of vortex nucleation during the process. We therefore check whether a moving pinning potential can generate a K\'arm\'an vortex street while dragging the initially pinned vortex, providing the $N_v=1$ analogue of the $N_v=0$ study of \citet{SasakiKarmanGP}, performed with~PBC.

We consider two simulations with different moving potentials. In the first scenario, we define an external potential as in Eq.~\eqref{eq:GaussPeak}. In the second, we instead use vertical Gaussian rods,
\begin{equation}
	\label{eq:GaussRod}
	V(\boldsymbol{x}) = \sum_{n \in \mathbb{Z}} V_0 \exp\left( -\frac{|x - x_n|^2}{2 \sigma^2} \right) \,.
\end{equation}
In both cases, the potential moves as $\boldsymbol{x}_p(t) = (x_{p,0} + v_p(t)\, t, ~ y_{p,0})$, where $p$ runs over all the peaks or rods, and the background velocity is defined as
\begin{equation}
	v_p (t) =
	\begin{cases}
		0 & t < 0 \\
		v_0 \left[ 1 - \cos\left(\frac{\pi \, t}{2 \, T}\right) \right] & 0 \le t \le T \\
		v_0 \left[ 1 + \frac{\pi}{2T} \left(t-T\right) \right] & t > T
	\end{cases}
	\label{eq:vBg}
\end{equation}
We choose $v_0 = c_1$, $T/\tau = 1 \times 10^4 \gg 1$, and run with a dissipation parameter $\gamma = 0.01$.
The velocity is increased adiabatically in time, slowly enough that dissipation can be assumed to equilibrate the system (and thus the superfluid loses memory of the evolution of $v_p$) at every moment~\citep{AntoHaskell}.

The simulations are performed on a $180\,\xi \times 60\,\xi$ cell with resolution $\delta l = 0.33\,\xi$.
The runs are initialized with a single vortex at the maximum of the Gaussian potential, for which we fix $V_0 = \mu$ and consider different values of~$\sigma / \xi = 0.5, 1, 2, 5$.
We start the runs at $t=-200\,\tau$ to dissipate spurious effects arising from the relaxation of the initial condition. This burn-in time is chosen to be long enough to obtain a numerically equilibrated initial condition at~$t=0$.

\subsubsection{Vortex depinning, nucleation and wake ($N_v=1$)}

\begin{figure*}[t]
	\centering 
	\hspace*{-8pt}\includegraphics[width=0.98\textwidth]{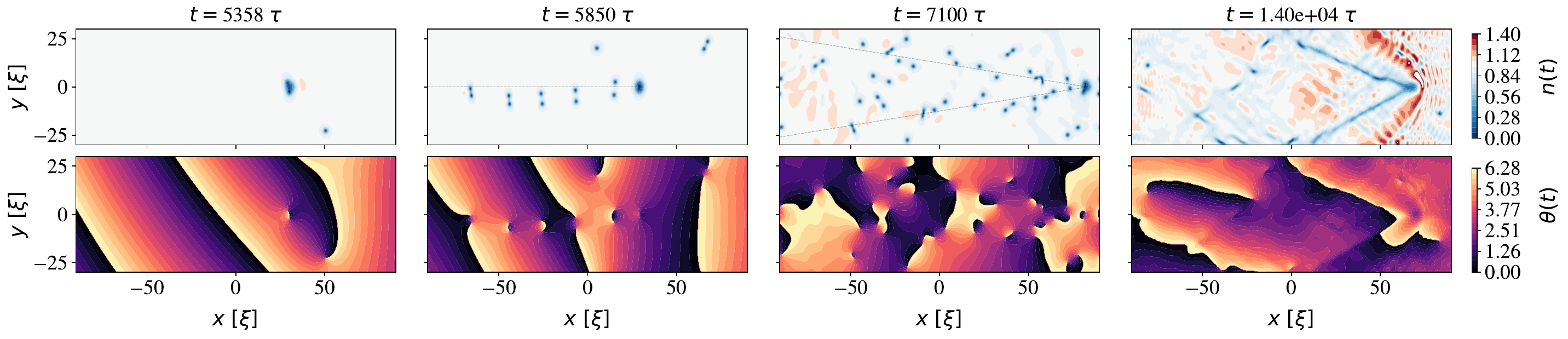}
	\caption{Evolution of the number density (\textit{top row}) and phase (\textit{bottom row}) for a superfluid with $N_v = 1$ in a moving array of Gaussian peaks. 
		The sequence shows the nucleation of the first vortex pair, the formation of single- and double-branched K\'arm\'an vortex streets, and a late stage where the potential, moving at $v_p > c_1$, generates a sonic boom. 
		Dashed black lines in the second and third panels indicate the approximate inclination of the K\'arm\'an vortex street. 
		Vortices and antivortices are distinguished by the orientation of the phase winding around points of complete depletion.}
	\label{fig:karmanStreet}
	\includegraphics[width=\textwidth]{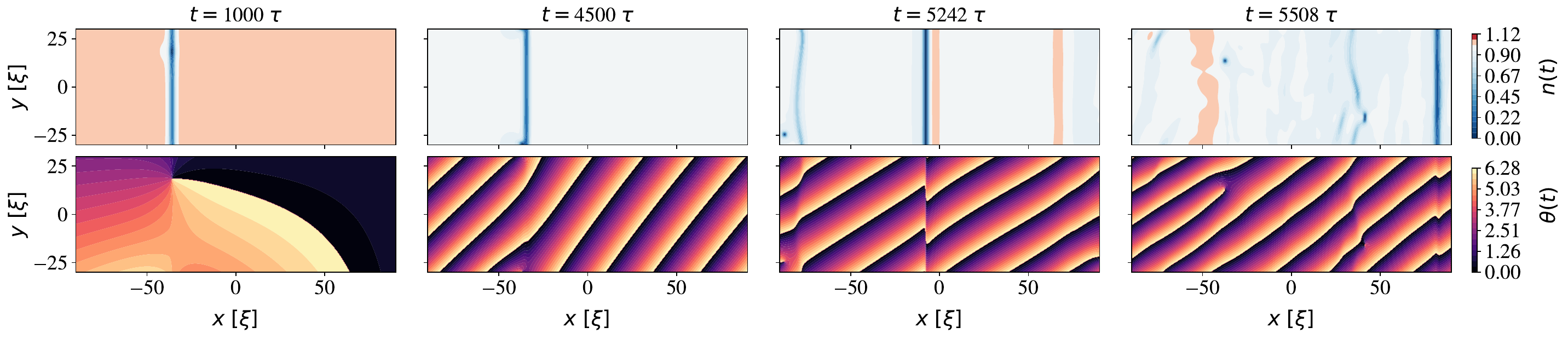}
	\caption{Evolution of the number density (\textit{top row}) and phase (\textit{bottom row}) for a superfluid with $N_v = 1$ immersed in a moving array of Gaussian rods. The system is initialized with a single vortex at the center of the background potential, at $y=0$. The sequence shows the state before depinning, the depinning of the original vortex, its encounter with the first soliton, and the nucleation of a vortex–antivortex pair induced by a snake instability of the soliton at~$x \simeq 45\,\xi$.}
	\label{fig:karmanRod}
\end{figure*}

In Fig.~\ref{fig:karmanStreet}, we show a series of snapshots from the Gaussian peak scenario with $N_v=1$ and $\sigma = 2$. At $t \simeq 5358\,\tau$, the background velocity has already exceeded the critical depinning threshold $v_\text{dep}$ and has just reached the nucleation threshold $v_\text{nucl}$.
Initially, the original vortex moves in the direction predicted by the Magnus-force term of the point vortex model.
The motion is independent from the prescription in Eq.~\eqref{eq:vBg}, and it is consistent with previous GPE~\citep{Liu2024} and point-vortex~\citep{AntoHaskell} simulations.
Its trajectory is perturbed only when it collides with the potential, interacting at distances of order $\sigma$.

Once $v_p \sim v_\text{nucl}$, the system begins to generate vortex–antivortex pairs\footnote{
    These pairs can be viewed as the two-dimensional analogue of the vortex loops nucleated in the numerical experiments of \citep{pecak2024toolkit}, since a vortex ring intersected by a plane parallel to its axis appears as a vortex-antivortex pair.}
from the moving peak, as imposed by the topological constraint $N_v=1$ and the Helmholtz theorem. 
These pairs are nucleated at quasi-regular time intervals, forming a K\'arm\'an vortex street, clearly visible in the second snapshot of Fig.~\ref{fig:karmanStreet}. 
For $N_v=0$, the vortex street would follow the potential along a straight line~\citep{SasakiKarmanGP}. In our simulation, however, the velocity field induced by the original vortex (at the top center of the fundamental cell) drags the pairs away from the $x$-axis.

As the background velocity increases, nucleation becomes more efficient, with vortex–antivortex pairs approaching each other, annihilating, and producing sound waves.  
As the number of pairs grows with $v_p$, the K\'arm\'an vortex street splits into two branches, visible at $t \simeq 7100\,\tau$ in Fig.~\ref{fig:karmanStreet}.
At this stage, the original vortex that was present at $t=0$ is essentially impossible to locate, and its influence on the street's shape becomes negligible. Vortices nucleated from neighboring peaks also enter the numerical domain and contribute to the deformation of the vortex street.

Further increasing the background velocity enlarges both the separation angle between the two branches of the wake, and the frequency of pair production. 
When $v_p \gtrsim c_1$, the nucleation frequency reaches its asymptotic limit: pairs form and annihilate almost immediately, while the opening angle of the K\'arm\'an street also saturates. 
Meanwhile, the potential, moving faster than the first sound, generates a high-density shock wave in front of it, while the continuous annihilation of pairs near the peak produces a cone of depletion reminiscent of a sonic boom. 
At these late times, most vortex pairs are annihilated, leaving once again a single unpaired vortex, clearly visible as a branch cut in the plot of the phase at $t\simeq1.4\times10^{4}~\tau$.

As a note, we observe that interactions between the pre-existing (already depinned) vortex and the background potential can lower the critical threshold for nucleation, leading to premature and extemporaneous vortex pairs, rather than systematic nucleation along a K\'arm\'an street. 
This phenomenon resembles the kick mechanism for depinning described in \citet{Warszawski_GPE_2011, Warszawski_unpinning, Warszawski_knock_2013}, as both arise from the increased relative velocity between the superfluid and the background potential caused by the local velocity field of the vortex.

%
Figure~\ref{fig:karmanRod} shows the evolution of the analogous system with the background potential of Eq.~\eqref{eq:GaussRod} and $\sigma = \xi$.
For $v_p < v_\text{dep}$, the vortex moves inside the rod toward the positive $y$-direction, as predicted by the Magnus-like term of the point vortex model~\citep{AntoHaskell}. 
At $t\simeq4500~\tau$, it leaves the pinning site and moves independently of Eq.~\eqref{eq:vBg}. At later times, the rod produces solitons, whose depletion decreases with increasing width of the potential.
Dissipation makes solitons unstable, leading to snake instabilities and subsequent vortex–antivortex nucleation, as described in \citet{Huang2003, Verma2017, Gaidoukov2021}. This process is accelerated by nearby vortices, whose local velocity fields deform the solitons.
As noted earlier, the interaction of the original vortex with the potential can lower the nucleation threshold by locally increasing the background velocity. 
With Gaussian rods, this effect is hard to avoid: at depinning, the vortex is expelled mainly in the vertical direction and remains close to the potential for a long time, unlike in the case of localized Gaussian peaks.

\subsubsection{Energetics and critical velocities for depinning~($N_v=1$)}

\begin{figure}[t]
	\centering 
	\hspace*{-5pt}\includegraphics[width=0.5\textwidth]{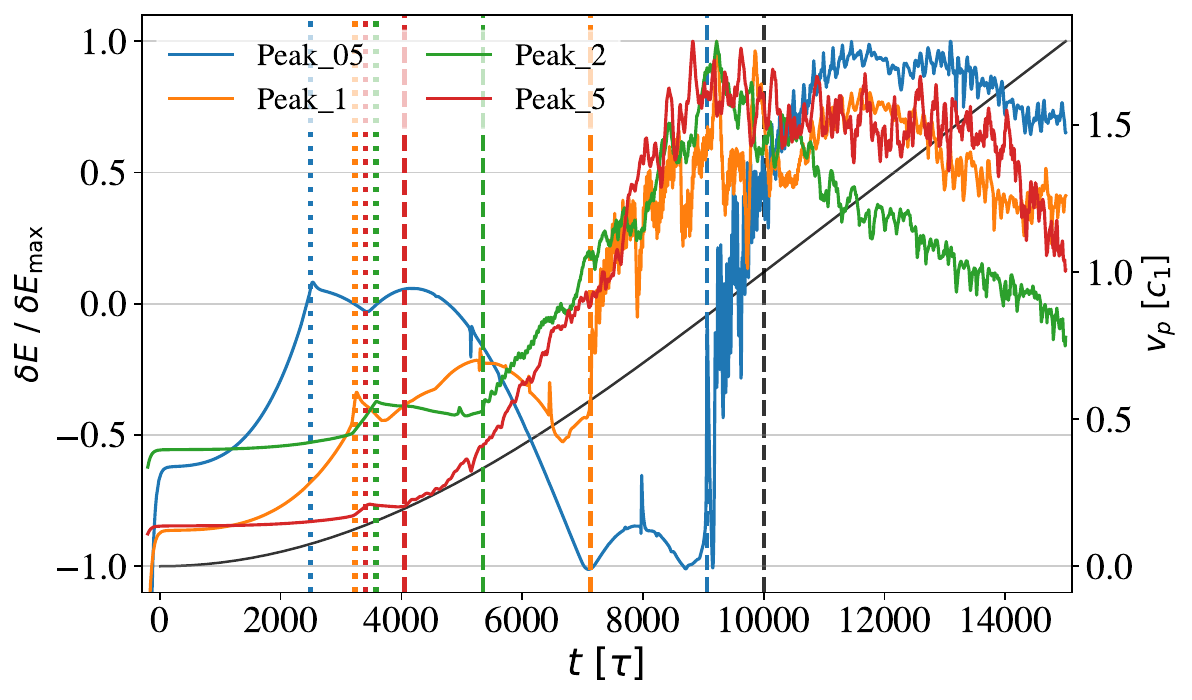}
	\caption{Evolution of the total energy $\delta E = E_\text{tot} - \langle E_\text{tot} \rangle$ for simulations with moving Gaussian peaks of widths $\sigma / \xi = 0.5, 1, 2, 5$. Colored dotted and dashed lines indicate the times of depinning and nucleation. The black solid and dashed lines show, respectively, the background velocity and the time it reaches $v_p = c_1$.}
	\label{fig:karman_energy}
\end{figure}

The evolution of the total energy is shown in Fig.~\ref{fig:karman_energy}.  
Negative times correspond to the burn-in phase, with fixed external potentials.
At $t=0$, the total energy of the superfluid has converged, confirming the thermalization of the initial condition.  
A first energy peak occurs at depinning. Once nucleation begins, if $\sigma \gtrsim \xi$, the total energy becomes roughly proportional to the number of vortex–antivortex pairs in the system, which act as unstable (due to dissipation) excitations of the velocity field. The energy rises at $v_p \sim v_\text{nucl}$ and decreases again at higher background velocities, when annihilation of pairs becomes more efficient than their nucleation.  

For $\sigma = 0.5\,\xi$, the potential is too narrow to consistently nucleate vortex–antivortex pairs. The two vortices form so close to each other, and already at high background velocities, that they almost immediately annihilate. In this case, the energy increase is due to the sound waves produced by such annihilations.  

Localized spikes in the energy evolution (\eg~at $t\sim5000~\tau$ in the $\sigma/\xi=0.5,2$ cases, or $t \sim 8000~\tau$ for $\sigma=0.5\,\xi$) indicate collisions between the original vortex and the moving potential. Such events can trigger (\eg~for $\sigma = 2\,\xi$) or fail to trigger (\eg~for $\sigma = 0.5\,\xi$) premature, extemporaneous nucleation depending on how close $v_p$ is to $v_\text{nucl}$.

Figure~\ref{fig:karman_energy} also shows the depinning time of the first vortex and the nucleation time of the first vortex–antivortex pair.  
Table~\ref{tab:crit_vel} summarizes the results, also including the simulations with Gaussian rods and their critical velocities for soliton production.
In the Gaussian peaks case, the critical velocity $v_\text{nucl}$ decreases monotonically with increasing potential width, while $v_\text{dep}$ instead saturates for $\sigma \gtrsim \xi$.
In the simulations with rods, the behavior of $v_\text{nucl}$ is complicated by extra triggers for nucleation, such as vortex-potential interactions and the decay of soliton snake instabilities.
For all potential widths and configurations (peaks and rods) tested here, depinning always precedes nucleation, although the gap between $v_\text{dep}$ and $v_\text{nucl}$ shrinks as $\sigma$ increases. Additional simulations with wider peaks are needed to explore the possible transition to $v_\text{nucl} \sim v_\text{dep}$. Finally, the markedly different values of $v_\text{dep}$ and $v_\text{nucl}$ between peaks and rods highlight the strong dependence of the critical parameters on the shape of the background potential.
\begin{table}[t]
	\centering
	\hspace*{-30pt}\begin{tabular}{ccccc}        
		\hline
		Potential & $\sigma~[\xi]$ & $v_\text{dep}~[c_1]$ & $v_\text{nucl} ~[c_1]$ & $v_\text{sol} ~[c_1]$ \\
		\hline
		\multirow{4}{*}{Peaks} & $0.5$ & $0.076$ & $0.853$ & - \\
		& $1$ & $0.126$ & $0.564$ & - \\
		& $2$ & $0.154$ & $0.334$ & - \\
		& $5$ & $0.139$ & $0.196$ & - \\
		\hline
		\multirow{4}{*}{Rods} & $0.5$ & $0.160$ & ($0.296$) & $0.329$ \\
		& $1$ & $0.242$ & $[0.350]$ & $0.312$ \\
		& $2$ & $0.291$ & $0.322$ & $0.369$ \\
		& $5$ & $0.243$ & $(0.258)$ & $0.328$ \\
		\hline
	\end{tabular}
	\caption{Critical velocities for depinning $v_\text{dep}$, nucleation $v_\text{nucl}$, and soliton creation $v_\text{sol}$ for two configurations of the moving external potential (Gaussian peaks and rods) and for different widths $\sigma$. Solitons appear only for the rod potential in Eq.~\eqref{eq:GaussRod}. Values in round and square brackets indicate cases where nucleation is triggered, respectively, by the interaction of the potential with the original vortex or by soliton snake instabilities.}
	\label{tab:crit_vel}
\end{table}


\section{Conclusions}
\label{sec:conclusions}

In this paper, we studied the dynamics defined by the Gross--Pitaevskii equation (GPE) in rectangular domains with the quasi-periodic boundary conditions (QPBC) in Eq.~\eqref{eq:prePbc}, first introduced (for static problems) in \citet{Mingarelli_2016,WoodTobyS.2019Qbcf}.
To do so, we implemented and tested \svapo, a C++ code that, unlike standard simulations with hard walls, harmonic potentials, or imposed zero net-vorticity, allows us to study the long-time evolution of the vortex dynamics in the bulk of a rotating superfluid. 
Unlike standard simulations based on hard walls or ordinary periodic boundaries, the QPBC allow us to study fluid elements with non-zero net vorticity, enabling the long-time tracking of vortex dynamics (a vortex never leaves the system) in a setting that resembles the genuine bulk limit (due to the absence of hard walls or confining potentials).

Although the GPE has been extensively studied in the literature for decades, the QPBC setup has received comparatively little attention. 
For this reason, we devoted significant effort to validating our code against analytical benchmarks and point-vortex predictions.
Tests that may appear elementary in more conventional settings (such as the stationarity of isolated vortices, the dynamics of vortex pairs, or relaxation toward vortex lattices) are less trivial in the presence of topological constraints associated with $N_v \neq 0$ and are therefore collected in Appendix~\ref{app:numChecks}. 
These tests are fundamental for validating both the theoretical and numerical framework, as simulations with QPBC are still not widespread (with the notable exceptions of the depinning simulations in~\citet{Liu2024} and the vortex-lattice formation simulations of~\citet{Doran2020}).

Building on these foundations, we investigated lattice formation (the natural case study for QPBC) in both homogeneous and structured environments.
As expected, dissipation drives the system toward Abrikosov order in the absence of external potentials, but produces frustrated configurations when pinning sites are present.

Finally, we examined vortex depinning and nucleation in moving potentials, showing the formation of K\'arm\'an vortex streets in systems with non-zero net vorticity.
This work thereby extends previous studies performed with $N_v=0$ and ordinary PBC (see \citep{SasakiKarmanGP}) to the topologically distinct case of~$N_v=1$.

The GPE on domains with QPBC can also serve as an explorative toy model for fluid elements in the bulk of neutron star crusts, allowing us to study relevant vortex processes as in \citet{Warszawski_knock_2013}, but without the limitations of a system confined in a trap. 
In this perspective, future developments may also include a systematic exploration of depinning thresholds for different pinning potentials (\eg~random pinning landscapes~\citep{AntoHaskell,link2022ApJ}), and extract mutual friction models for pulsar glitches~\citep{amp_review_2023}.

More generally, in both neutron-star and laboratory settings, the use of QPBC per se does not introduce new physics (and, as shown by our tests, yields results that are largely consistent with expectations). Rather, it should be regarded primarily as a tool to investigate vortex dynamics and superfluid turbulence under global topological constraints that cannot be captured using ordinary PBC.
As an example of a future application, it would be interesting to revisit the study of the pinning–unpinning transition for point vortices presented in \citep{AntoHaskell} within a GPE setting, which is better suited for such simulations, as it naturally incorporates vortex-core physics and acoustic radiation.

\begin{acknowledgments}
	FM acknowledges support from the Deutsche Forschungsgemeinschaft
	(DFG) under Grant No. 406116891 within the Research Training Group
	RTG 2522/1.
	MA acknowledges partial support from the IN2P3 Master Project NewMAC, and the ANR project GW-HNS, ANR-22-CE31-0001-01.
	The authors would like to thank Toby Wood and Daniel P\k{e}cak for insightful feedback and helpful discussions.
\end{acknowledgments}

\bibliographystyle{apsrev4-2}

\appendix

\section{Consistency checks}
\label{app:numChecks}

In this appendix, we present a series of more fundamental numerical experiments to validate our framework and code.
Again, Eq.~\eqref{eq:origGPE} is solved with the time-dependent QPBC of Eq.~\eqref{eq:prePbc}.
The real and imaginary parts of $\psi$ are evolved with a third-order Adams–Bashforth scheme~\citep{Butcher:2016}, while spatial derivatives are evaluated using five-point finite-difference stencils.
We adopt the natural units of Sec.~\ref{subsec:gpe}. Unless otherwise stated, we set $n_\infty=\mu=1$, the time step is $\delta t = 5 \times 10^{-4}\,\tau$ (always satisfying the CFL condition for the GPE), and work in the non-dissipative regime,~$\gamma=0$.

\subsection{Soliton waves}
\label{subsec:soliton}

As a first consistency check, we test our code against soliton analytical solutions. In an infinite domain, the field configuration
\begin{equation}
	\label{eq:soliton}
	\psi_S (\boldsymbol{x}, t) = \left[ i \sqrt{n_\textrm{min}} + \sqrt{n_\infty - n_\textrm{min}} \, \tanh\!\left(\frac{x - ut}{\sqrt{2} \,\xi_u} \right) \right] e^{-i\mu t}
\end{equation}
describes a soliton propagating with velocity
\begin{equation}
	\label{eq:soliton_vel}
	u = \sqrt{\frac{n_\textrm{min}}{n_\infty}} \, c_1 = \sqrt{\frac{\mu\, n_\textrm{min}}{n_\infty}} \,,
\end{equation}
where $n_\textrm{min}$ is the minimum density associated with the soliton depletion and $\xi_u \equiv \xi \left(1 - n_\textrm{min}/n_\infty \right)^{-1/2}$~\citep{pethick_book_2008}.

We consider a finite domain with zero net vorticity, for which the QPBC reduce to ordinary PBC.
A single soliton (which solves the GPE only in an infinite domain) is not compatible with PBC, as it would introduce phase a discontinuity at the cell boundaries.
By contrast, the combined wave function
\begin{equation}
	\label{eq:sol_init}
	\psi_{2S}(\boldsymbol{x}, t) = \psi_S(\boldsymbol{x}, t)\,\psi_S(-\boldsymbol{x}, t)    
\end{equation}
satisfies PBC and can therefore be simulated using our code.\footnote{
    A more elegant and rigorous way of imprinting solitons under PBC would be to use the exact solutions expressed in terms of Jacobi elliptic functions; see, for example, \citep{Tsuzuki_1971,carr_PRA_2000}.
    The simple ansatz \eqref{eq:sol_init} already serves our purposes.}
It describes the evolution of a pair of solitons propagating in opposite directions.

\begin{figure}[t]
	\centering
	\hspace{-5pt}\includegraphics[width=0.49\textwidth]{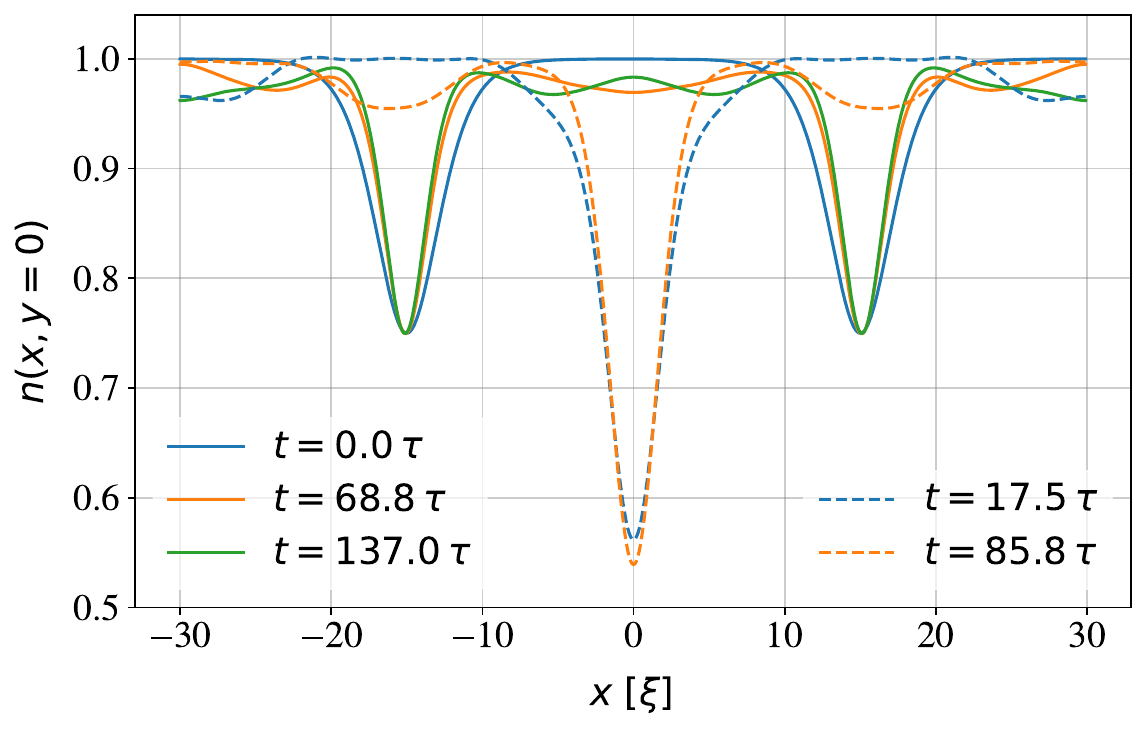}
	\caption{Superfluid density along the $x$-axis for some selected times for the two-solitons test with $n_\text{min} = 0.75$. Solid [dashed] lines represent a series of configurations with the solitons in $x = \pm 15\,\xi$ [$x \simeq 0$] and moving towards [away from] $x=0$.}
	\label{fig:solitonDens}
\end{figure}

The superfluid is initialized with a pair of solitons $\psi_{2S}$ and evolved up to $t_\text{max} = 150\,\tau$ in a rectangular fundamental cell of size $60\,\xi \times 30\,\xi$ with spatial resolution $\delta l = 0.1\,\xi$. We explore grey solitons with different values of their maximum depletion ($n_\text{min} = 0.2, 0.5, 0.75$), as well as dark solitons ($n_\text{min} = 0$).
In Fig.~\ref{fig:solitonDens}, we show the number density of the superfluid along the $x$-axis at different times. As expected, the two solitons propagate in opposite directions and cross each other multiple times.

The initial condition in Eq.~\eqref{eq:soliton} is exact only for a single soliton in an infinite system. Therefore, there is an initial transient relaxation during which secondary solitons with smaller depletion are generated, which travel faster than the imprinted pair. Within a few $\tau$, the imprinted field configuration stabilizes, and the resulting soliton shows a narrower depletion than the analytical prediction. The solitons cross the cell boundaries and reproduce the same qualitative profile every $\Delta t \simeq L/u$. Quantitative differences in the density profile at different times, as well as its oscillations, are mainly due to the propagation of the secondary solitons, which are only a by-product of our initial ansatz in Eq.~\eqref{eq:sol_init}. 

In our simulations, we do not observe the snake instabilities reported \eg~in \citet{Huang2003, Verma2017, Gaidoukov2021}. Including a finite dissipation $\gamma$ only damps the wave by gradually reducing the depletion with time, as expected.

\begin{figure}[t]
	\centering
	\hspace{-5pt}\includegraphics[width=0.47\textwidth]{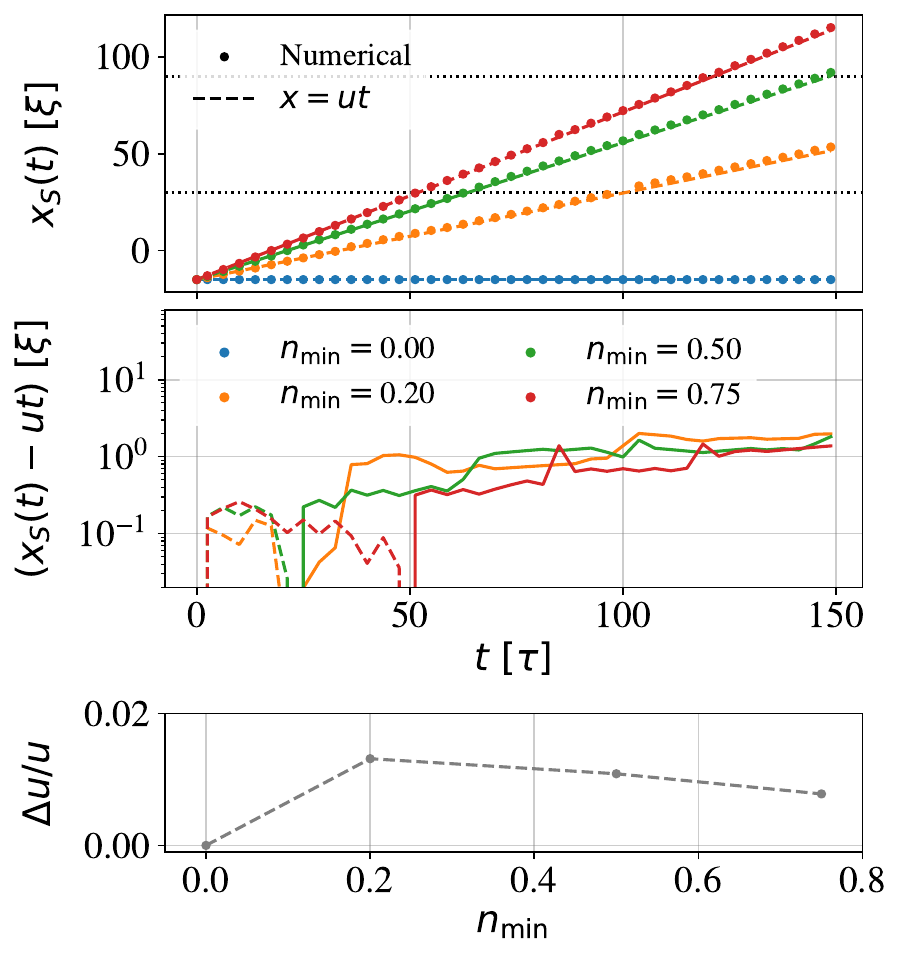}
	\caption{Tracking of the soliton moving to the right in the two-solitons test for different values of $n_\text{min}$.
		\textit{Top}: Comparison between the position of the soliton as analytically calculated from Eq.~\eqref{eq:soliton_vel} and as tracked as the moving maximum depletion. The horizontal, dotted lines show the $x$-boundaries of the periodic fundamental cell.
		\textit{Central}: Absolute difference between the numerical and analytical predictions for the position of the soliton. The dashed lines represent negative values.
		\textit{Bottom}: Relative discrepancy between the numerical and analytical soliton velocities as a function of the initial maximum depletion associated with each soliton.}
	\label{fig:solitonVel}
\end{figure}

\begin{figure}[t]
	\centering 
	\hspace{-10pt}\includegraphics[width=0.48\textwidth]{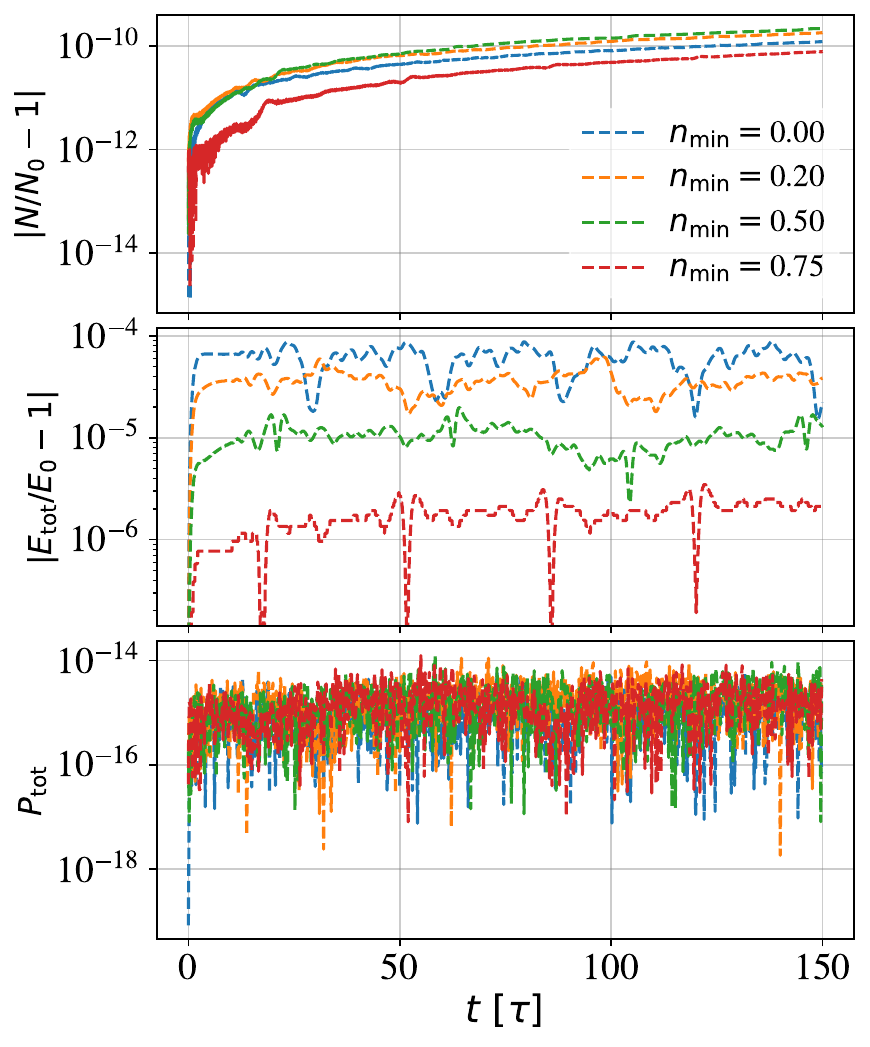}
	\caption{Relative evolution of the total number of particles, total energy and total momentum of the superfluid for the two-solitons tests.}
	\label{fig:solitonCons}
\end{figure}

We check that the solitons move with the analytical velocity predicted in Eq.~\eqref{eq:soliton_vel}, see Fig.~\ref{fig:solitonVel}. The few-per-cent error is due to the small discrepancy between the numerical structure of the solitons and the analytical result of Eq.~\eqref{eq:soliton}. 
The evolution of the total number of particles, energy, and momentum is shown in Fig.~\ref{fig:solitonCons}. 
As expected, these quantities are conserved during the evolution of the system. At the beginning of the simulation, a burn-in phase lasting a few dynamical timescales occurs as the system relaxes from the approximate initial configuration.

\subsection{Simple vortex configurations}
\label{sec:simpleVortConfigs}

In this section, we further test the stability and reliability of our code by running simulations of simple vortex configurations and comparing the results with the predictions of the point vortex model \cite[\eg][]{AntoHaskell,lucas2014,skipp2023}.

\subsubsection{Single vortex and sound waves}
\label{subsec:singleVandSoundWaves}

In large rotating containers, vortices arranged in a simple Bravais lattice form a stable configuration that rotates uniformly with constant angular velocity. 
We test here that our implementation of the QPBC is consistent with this property: a vortex imprinted at any position in our rectangular domain (in the absence of an external potential, $V=0$) must remain stationary. 
Since our aim is to develop a tool for simulating fluid elements that reproduce the features of bulk behavior, this is also a desirable outcome, consistent with the fact that a single vortex in an infinite plane (or far enough from walls) remains stationary unless additional currents are introduced.

\begin{figure}[t]
	\centering 
	\hspace*{-10pt}\includegraphics[width=0.49\textwidth]{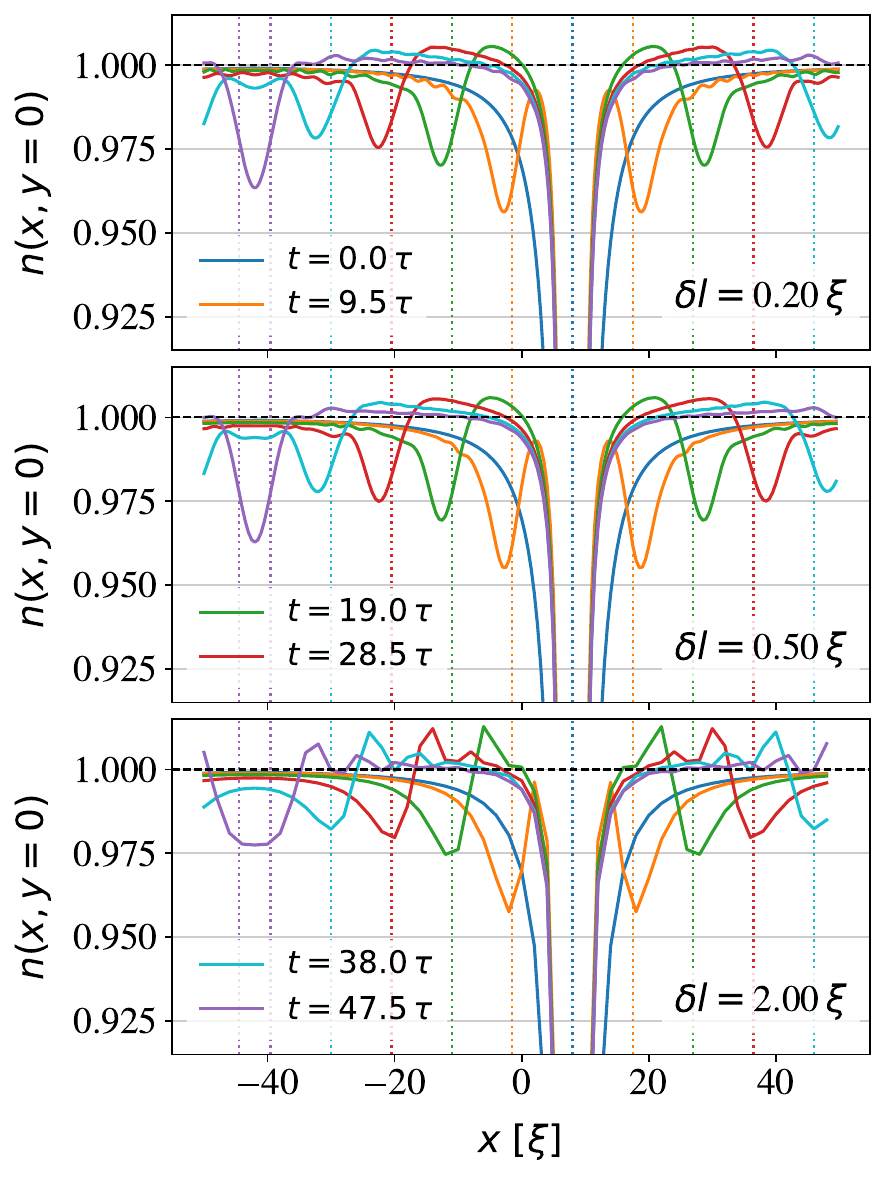}
	\caption{Evolution of the density when a single vortex is imprinted at $t=0$ (corresponding to the dark blue curve) in $\boldsymbol{x} = (8\,\xi,0)$. 
		Different panels correspond to different grid resolutions.
		The black dashed lines indicate the bulk density $n_\infty=1$. For comparison, the vertical, dotted lines show the predicted position of the wave fronts assuming that they travel at the constant speed~$c_1$.}
	\label{fig:freeVortexDens}
\end{figure}

We test this behavior by initializing a superfluid with a singly charged vortex at $\boldsymbol{x} = (8\,\xi,0)$. We use a square cell with $L = 100\,\xi$ and consider a series of spatial resolutions $\delta l/\xi = 0.2, 0.5, 2$. 
Since the initial condition in Eq.~\eqref{eq:init_cond} only approximates the depletion in the vortex core, we expect the system to relax to a more stable configuration over a few dynamical timescales by generating radial sound waves from the vortex center (vortex breathing). On longer timescales, these sound waves are damped by numerical dissipation.
Figure~\ref{fig:freeVortexDens} shows the evolution of the superfluid density along the $x$-axis (the system is spherically symmetric around the vortex center). 
As expected, the vortex position remains stable indefinitely, independent of the grid resolution or the initial vortex position. 
The vortex breathing wave propagates through the superfluid with a velocity slightly faster than $c_1 = \sqrt{\mu}$.

\subsubsection{Vortex couples} 
\label{subsec:vCouples}

To check how the QPBC setting deviates from a genuine bulk limit, we compare our numerical results with the orbits of point vortices in the infinite plane. In this limit, the modulus of the velocity of the two point vortices is $v_{1,2} = s_{2,1}/d$, where $d$ is their mutual distance\footnote{
	Since the point vortex approximation neglects compressibility and finite core size, it is valid when the inter-vortex separation and other scales of the system are much larger than $\xi$ (see \citep{lucas2014,skipp2023} for a revised derivation of point-vortex dynamics from the GPE). In our case, this translates into the requirement $d, L \gg \xi$, while the genuine bulk limit is achieved for~$L\gg d$.}.
It is well known that $d$ does not change if there is no dissipation: two equally charged vortices orbit around each other (anticlockwise for positive charges), while two vortices with opposite orientations move along a straight line (in our case, in the positive $y$-direction).

We consider two sets of simulations with two vortices, either with $N_v = 0$ (vortex–antivortex in PBC) or $N_v=2$ (two vortices with the same orientation in QPBC).
We start the runs with two vortices in $\boldsymbol{x}_{1,2} = (\pm x_0, 0)$ at $t=0$, with $x_0/\xi = 2, 3, 4, 5, 6$ (namely, $d=2x_0$), and $s_{1,2} = \mp 1$ or $s_{1,2} = 1$ in the two cases.
All simulations are performed in a square cell with $L = 100\,\xi$ and spatial resolution $\delta l = 0.2\,\xi$.
We carry out convergence tests on the $x_0 = 2$, $N_v = 2$ case also considering the resolutions~$\delta l/\xi = 0.1, 0.4, 0.8$. 

\begin{figure}[t]
	\centering 
	\hspace{-10pt}\includegraphics[width=0.5\textwidth]{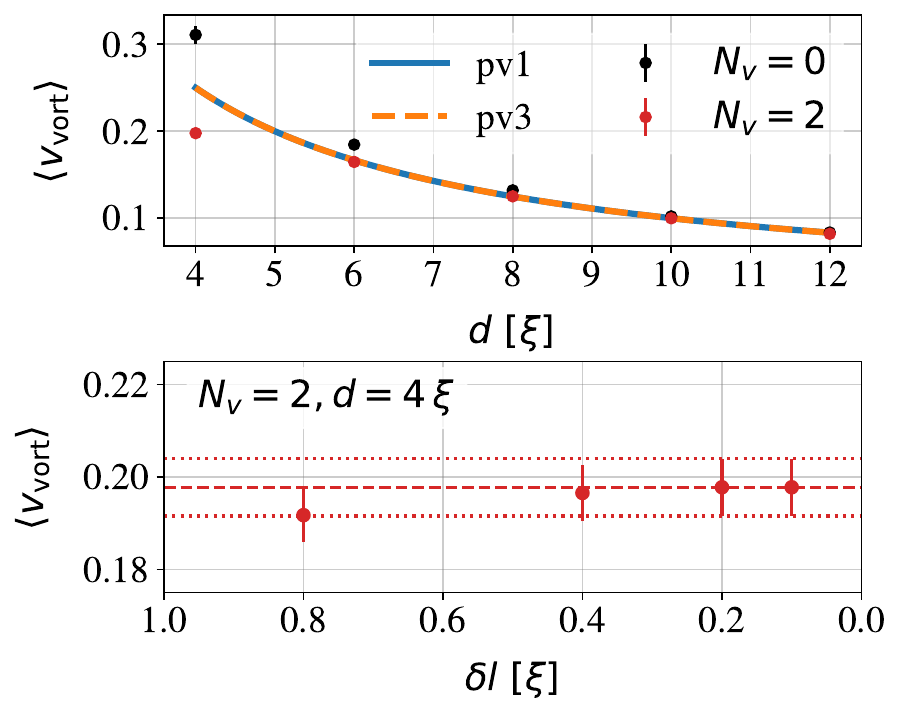}
	\caption{
		\textit{Top}: Vortex velocity as a function of vortex separation for the aligned ($N_v = 2$, red dots) and anti-aligned ($N_v = 0$, black dots) cases. The blue solid line and the orange dashed line show the point vortex model predictions obtained by considering only the vortices in the fundamental cell, or those in a $3\times 3$ array of cells centered on it, respectively. \textit{Bottom}: Vortex velocity for the aligned case with $d = 4\,\xi$ as a function of grid resolution. The dashed and dotted red lines mark the estimate from the run with $\delta l = 0.2\,\xi$, also reported in the top panel.
	}\label{fig:vCoupleVel}
\end{figure}

To check that our simulations effectively capture the genuine bulk limit ($L \gg d$), we demand that the point vortex model gives the same prediction whether only the fundamental cell or also the eight surrounding image cells are included. This is shown in Fig.~\ref{fig:vCoupleVel}. 
Smaller cells would also introduce oscillations on top of the constant speed $v_{12}$ for the $N_v = 2$ case, as the separation between vortices of different image couples is not constant in time. 
The $N_v=0$ case does not show this effect, since the system evolves via rigid translation of the initial configuration and all image vortices.

The top panel of Fig.~\ref{fig:vCoupleVel} shows the observed average vortex speed $\langle v_\text{vort} \rangle$ compared with the constant $v_{12}$ coming from the point vortex model (see Appendix~\ref{app:vort_vel} for details on how $\langle v_\text{vort} \rangle$ is computed). 
As expected, the speed $\langle v_\text{vort} \rangle$ decreases with increasing vortex separation, and agrees with the analytical $v_{12}$ within $\sim1\%$ and $\sim5\%$ relative errors for $d \gtrsim 6\,\xi$ and $d \gtrsim 8\,\xi$ in the $N_v=2$ and $N_v=0$ cases, respectively. 
At smaller separations, where $d$ becomes comparable to the vortex core size (a few healing lengths), the dynamics is affected by the core structure. 
In addition, we find that the finite-size corrections have opposite signs in the aligned and anti-aligned cases.

To ensure that these effects are not spurious, the bottom panel of Fig.~\ref{fig:vCoupleVel} shows the convergence of the average vortex velocity for the aligned case with $d = 4\,\xi$, demonstrating that a resolution $\delta l < 0.4\,\xi$ is enough to provide an estimate of~$\langle v_\text{vort} \rangle$ accurate to the $<1\%$ relative error.

\begin{figure*}[t]
	\centering 
	\hspace{-5pt}\includegraphics[width=0.98\textwidth]{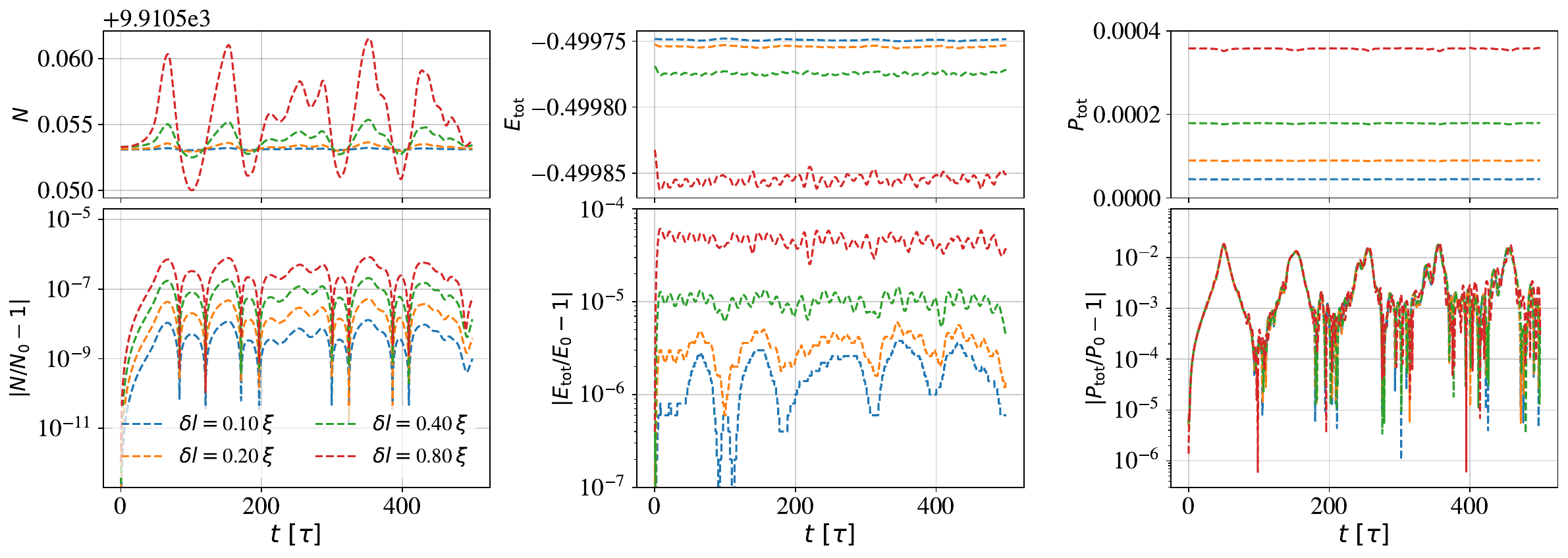}
	\caption{
		Evolution of the conserved quantities during the vortex-couple run with $d = 4\,\xi$ at different grid resolutions. \textit{Top}: Evolution of the total particle number, energy, and momentum of the superfluid. \textit{Bottom}: Relative difference between the conserved quantities and their initial values. The oscillations and noise in the total momentum arise from the propagation of vortex-breathing sound waves.
	}
	\label{fig:vCoupleCons}
\end{figure*}

In Fig.~\ref{fig:vCoupleCons} we show the conserved quantities for the case $N_v = 2$, $d = 4\,\xi$, at a series of different grid resolutions.
The initial values of the total number of particles, energy, and momentum converge rapidly for $\delta l < 0.4\,\xi$. 
The oscillations in the total momentum are due to sound waves generated by the initial vortex breathing. Their period corresponds to the time required for a wave front to propagate along a cell side with $c_1 = 1$ in code units. 
At late times, sound waves from increasingly distant periodic cells combine to produce noisy perturbations of the asymptotically stable configuration.

While the initial condition in Eq.~\eqref{eq:init_cond} correctly reproduces the particle number and total energy of the system (in the limit of infinite resolution), it reproduces the total momentum only within a relative error of order $10^{-3} - 10^{-2}$. 
The estimate holds only for this specific superfluid configuration. Equation~\eqref{eq:init_cond} performs differently in reproducing the total momentum for different setups (number and positions of vortices, as well as the dimensions and geometry of the cell).
An initially dissipative ($\gamma > 0$) evolution allows to thermalize the initial condition, and can thus improve convergence. Note however that a dissipative dynamics would change the inter-vortex distance with time.

\subsubsection{Single vortex in Gaussian potentials}
\label{subsec:gaussPeak}

To probe the coupling of the superfluid with an external potential $V$, we consider the basic test of a singly charged vortex in the presence of a Gaussian peak \cite[\eg][]{SasakiKarmanGP,link2009PhRvL,AntoHaskell,Liu2024}.
We consider a square cell with $L = 30\,\xi$ and set the spatial resolution to $\delta l = 0.2\,\xi$.
We fix the potential as in Eq.~\eqref{eq:GaussPeak}, impose $\sigma = 3\,\xi$, and run a set of simulations for different values of the potential strength $V_0$ and the initial vortex-peak distance $d$.
Both attractive ($V_0 < 0$) and repulsive ($V_0 > 0$) cases are considered.

\begin{figure*}[t]
	\centering 
	\includegraphics[width=\textwidth]{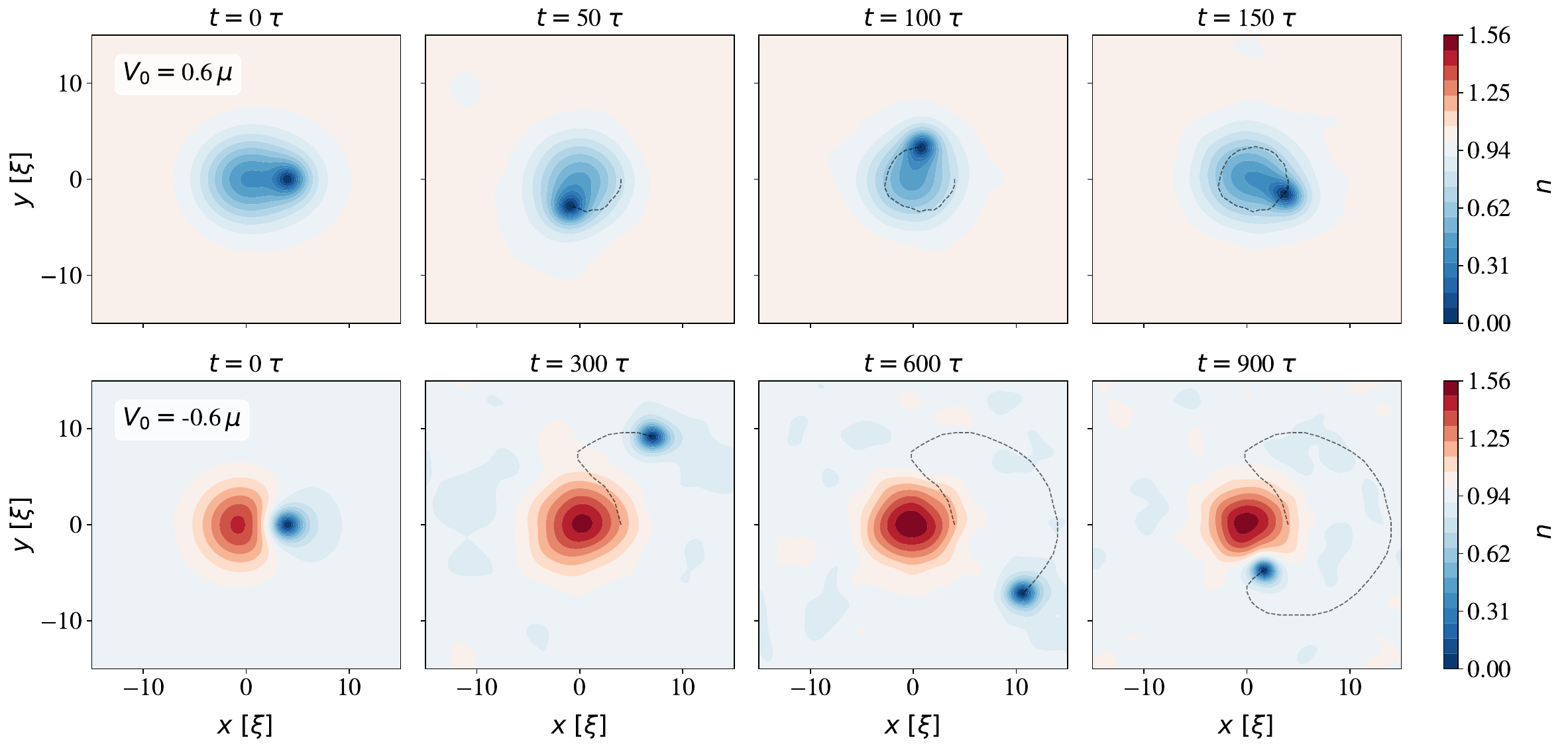}
	\caption{Selected snapshots of the superfluid density for a simple lattice of singly charged vortices in the periodic Gaussian potential defined in Eq.~\eqref{eq:GaussPeak}. 
		\textit{Top} [\textit{bottom}] \textit{row}: repulsive [attractive] potential with $|V_0| = 0.6\,\mu$ and initial vortex distance $d = 4\,\xi$. 
		Blue and red areas correspond to regions of depletion and accretion relative to the mean superfluid density $n_\infty = 1$. 
		The dashed black line shows the vortex trajectory.}
	\label{fig:gaussPeakDens}
\end{figure*}

Figure~\ref{fig:gaussPeakDens} shows snapshots for $V_0 = \pm 0.6\,\mu$ with $d = 4\,\xi$.
The superfluid is either repelled from or attracted toward the Gaussian peak, modifying the velocity field induced by the vortex.
In the repulsive case, the vortex rotates clockwise around the Gaussian peak (see also \citep{Liu2024}), while in the attractive case it follows a more complicated closed trajectory.
As already discussed, the vortex remains stationary in the quasi-periodic cell if $V=0$.
Unlike the point-vortex case\footnote{
	To model vortex motion in non-homogeneous media, the point-vortex approach introduces an effective potential that acts directly on the vortex position. 
	From the phenomenological equation of motion, it follows immediately that flipping the external potential simply reverses the velocity along the same circular orbit \cite[\eg][]{AntoHaskell}.},
the orbit does not simply reverse under $V_0 \rightarrow -V_0$.
This asymmetry arises because the potential $V$ in the GPE acts on the condensate density, inducing either a local depletion ($V_0>0$) or an enhancement ($V_0<0$). By contrast, the vortex core always produces complete density depletion, making the two cases asymmetric.

\begin{figure}[t]
	\centering
	\hspace{-10pt}\includegraphics[width=0.5\textwidth]{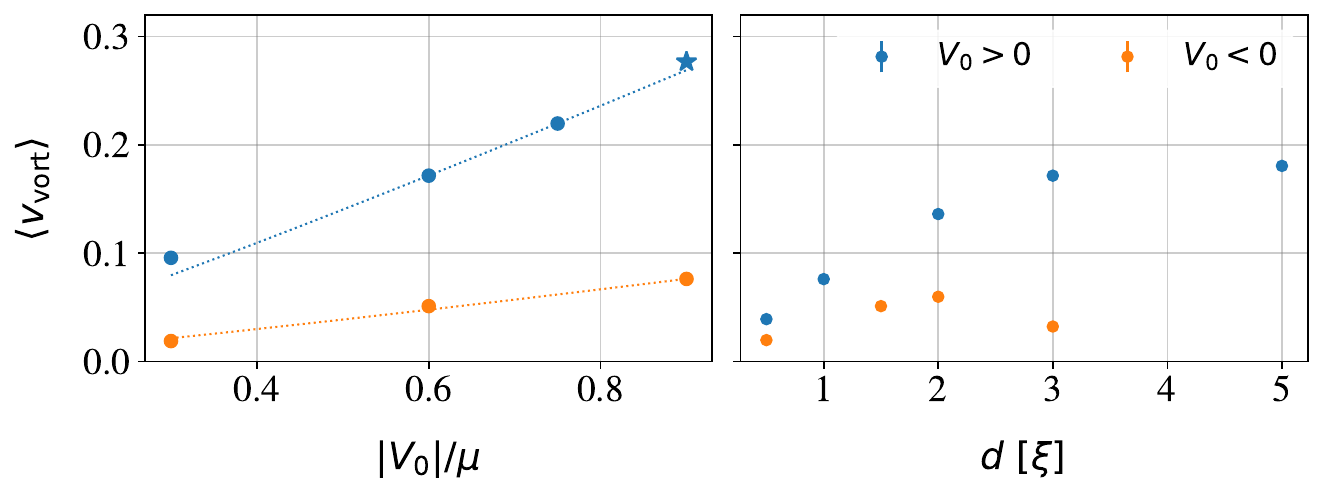}
	\caption{
		Vortex velocity during stable rotation around the central Gaussian peak defined in Eq.~\eqref{eq:GaussPeak} with width $\sigma = 3\,\xi$. 
		In blue and orange, the repulsive and attractive cases. 
		\textit{Left}: Vortex velocity as a function of the absolute value of the potential strength. The vortex distance from the center is $d/\xi = 1.5, 3$ for the attractive and repulsive cases, respectively. The star marks the only observed dissipative configuration, where the vortex falls into the depletion caused by the Gaussian peak. 
				\textit{Right}: Vortex velocity as a function of the distance from the Gaussian peak for fixed potential strength $V_0 = \pm 0.6\,\mu$.} 
	\label{fig:gaussPeakVel}
\end{figure}

The geometry of the fundamental cell also plays a role. 
In the attractive case of Fig.~\ref{fig:gaussPeakDens}, the vortex starts close enough to the cell boundary to be influenced by the square-like symmetry imposed by the fundamental cell (or equivalently, by the neighboring potential peaks). If the vortex is initialized closer to the center of the cell, its motion is instead governed by the local radial symmetry of the central Gaussian peak. 
For instance, setting the vortex at $d = 3\,\xi$ is sufficient to make it rotate anti-clockwise in a stable orbit around the center of the cell.

Finally, a vortex orbiting a Gaussian peak is expected to move faster if $|V_0|$ is increased.  
We test this in Fig.~\ref{fig:gaussPeakVel}, where we measure the average rotational velocity of the vortex around the Gaussian peak (see Appendix~\ref{app:vort_vel} for details). 
We consider a range of potential strengths, and distances $d/\xi = 1.5, 3$ for the attractive and repulsive cases, respectively.
We observe that a sufficiently strong repulsive potential can induce dissipation in the orbital motion: for $V_0 = 0.9\,\mu$, the vortex is no longer on a stable orbit but spirals into the potential peak\footnote{
	Note that $V_0 = \mu$ is the maximum observable potential in our configuration (with $\mu = 1$), since at this value the superfluid would be completely depleted.}.
The orbital velocity also depends on the width of the potential (see the right panel of Fig.~\ref{fig:gaussPeakVel}).
The orbits stops both in the limit $d=0$, for which the vortex is stable in the center of the external potential, or if the vortex is initialized close enough to the cell boundaries to get trapped by the combined effect of the central and of the neighboring Gaussian peaks.
The distance from the peak that maximizes $\langle v_\text{vort} \rangle$ depends on the sign of the potential.

\subsection{Kelvin circulation theorem}
\label{subsec:Kelvin}

For an inviscid, barotropic fluid subject to conservative body forces, the Kelvin circulation theorem states that the circulation
\begin{equation}
	\label{eq:KelvinCirc}
	\Gamma(t) = \oint_{C(t)} \boldsymbol{v} \cdot d\boldsymbol{l}
\end{equation}
around a closed material curve $C(t)$ is conserved,
\begin{equation}
	\label{eq:KelvinTheo}
	\frac{d\Gamma(t)}{dt} = 0 \, .
\end{equation}
This result is essentially equivalent to the local Helmholtz equation, $\partial_t \boldsymbol{\omega} = \nabla \times (\boldsymbol{v} \times \boldsymbol{\omega})$, with $\boldsymbol{\omega}=\nabla\times \boldsymbol{v}$ \cite[\eg][]{ArnoldKhesin}. 

\begin{figure*}[t]
	\centering 
	\includegraphics[width=\textwidth]{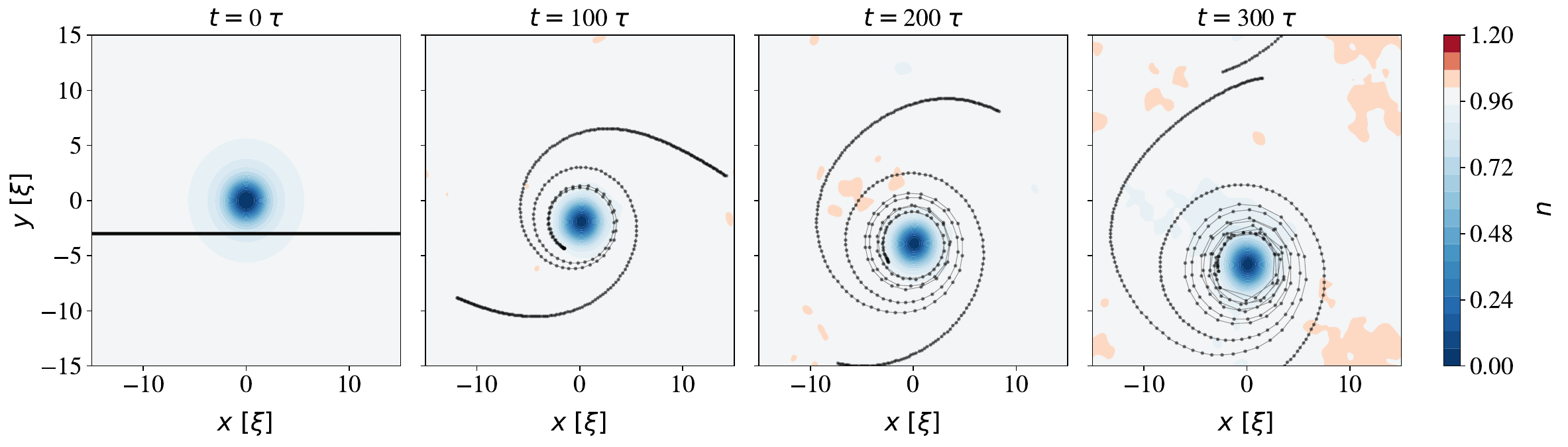}
	\caption{Snapshots of the superfluid density for a lattice of singly charged vortices dragged by the periodic Gaussian potential of Eq.~\eqref{eq:GaussPeak}. 
		Blue and red indicate regions of depletion and accretion relative to the mean density $n_\infty = 1$. 
		Black dots mark the tracers defining the closed (in QPBC) circuit used to test the Kelvin theorem~\eqref{eq:KelvinTheo}. 
		The circuit deforms and wraps around the vortex without crossing it, so the circulation is conserved.}
	\label{fig:kelvinDens}
\end{figure*}

In Appendix~\ref{app:vortEq_HelmTheo}, we show that although the Helmholtz equation does not apply to condensates governed by the GPE, the Kelvin circulation theorem remains valid.  
Hence, another interesting test is to see numerically the physical consequences of the theorem. To do so, we pin a vortex on a Gaussian peak and slowly move it with constant velocity $\boldsymbol{v}_p = -0.02\,c_1 \, \hat{\mbf{y}}$. We set $V_0/\mu = \sigma/\xi = 1$ in Eq.~\eqref{eq:GaussPeak} and consider a squared cell with side $L=30\,\xi$ and spatial resolution $\delta \simeq 0.083$.
A closed (in QPBC) circuit $C(t)$ is tracked by placing 360 tracers along $y = -3\,\xi$ and evolving them with the local velocity field. 
As shown in Fig.~\ref{fig:kelvinDens}, while the vortex moves downward, $C(t)$ deforms to avoid any crossing, and the circulation $\Gamma(t)$ remains constant, consistently with the Kelvin theorem.
This is indeed the expected behavior, as it is well known that, within the GPE framework, a vortex moves only in response to local derivatives of the wave function, even in the presence of an external potential~$V$~\citep{BiaBiaSliwa}. 

\section{Vorticity equation and Helmholtz theorem}
\label{app:vortEq_HelmTheo}

We expand on the analysis of \citep{dosSantos} to show the validity of the circulation theorem for a condensate described by a GPE with~$\nabla V \neq 0$ and $N_v\neq 0$.

We start from Eq.~\eqref{eq:origGPE}, and use the Madelung representation $\psi(\boldsymbol{x}, t) = \sqrt{n(\boldsymbol{x}, t)} \, \exp[i \theta(\boldsymbol{x}, t)]$, where the phase $\theta(\boldsymbol{x}, t)$ is a multivalued real field \citep{kleinert2008multivalued}. 
It follows the conservation law $\partial_\mu (\epsilon^{\mu \nu \alpha \beta} F_{\alpha \beta}) = 0$, where $\epsilon$ is the Levi-Civita symbol, and $F_{\alpha \beta}$ is the force field tensor defined in \citet{dosSantos}, namely $F_{\alpha \beta} = \partial_\mu v_\nu - \partial_\nu v_\mu$, with $v_\nu = (\psi^* \partial_\mu \psi - \psi \partial_\mu \psi^*)/(2i \psi^* \psi)$.

The space components $\partial_\mu (\epsilon^{\mu i \alpha \beta} F_{\alpha \beta}) = 0$ for $i=(x,y,z)$ give the vorticity transport equation
\begin{equation}
	\label{eq:vortTranspEq}
	\partial_t\, \mbf{\omega} + \nabla \times (\mbf{\omega} \times \boldsymbol{v}) = \nabla \times \mbf{f}^v \,,
\end{equation}
with $\mbf{f}^v = \partial_t \mbf{\theta}^v + \nabla \theta_t^v - \boldsymbol{v} \times (\nabla \times \boldsymbol{v})$, where $(-\theta_t^v/c, \mbf{\theta}^v)$ are the components of the vortex gauge field defined in \citet{dosSantos}.
They represent the force per unit mass acting on the condensate due to the presence of quantized vortices.

The total derivative of the circulation is given by
\begin{equation}
	\label{eq:circulation_der}
	\frac{d \Gamma(t)}{dt} = \oint_{C(t)} \frac{D \boldsymbol{v}}{Dt} \cdot d\boldsymbol{l} + \oint_{C(t)} \boldsymbol{v} \cdot \frac{D \boldsymbol{l}}{Dt} \,,
\end{equation}
where $D/Dt = \partial_t + \boldsymbol{v} \cdot \nabla$ is the Lagrangian derivative operator and $d\boldsymbol{l}$ is the oriented line element. The second term vanishes as in the classical case \cite[\eg][]{clarke_carswell_2007}. Using the Stokes theorem, one can rewrite Eq.~\eqref{eq:circulation_der} as
\begin{equation}
	\frac{d \Gamma(t)}{dt} = \int_{S(t)} \left[\partial_t \mbf{\omega} - \nabla \times \left( \boldsymbol{v} \times \mbf{\omega} \right) \right] \cdot d\mbf{a} \,,
\end{equation}
where $S(t)$ is the surface delimited by $C(t)$ and $d\mbf{a}$ is the surface element pointing outwards.
Inserting here Eq.~\eqref{eq:vortTranspEq} and using again the Stokes theorem leads to
\begin{equation}
	\frac{d \Gamma(t)}{dt} = \oint_{C(t)} \left( \nabla \times \mbf{f}^v \right) \cdot d\boldsymbol{l} \,.
\end{equation}
The integral vanishes as long as the closed line $C(t)$ does not (initially) intersect any vortex line, as the force $\mbf{f}^v$ exclusively acts at the vortex positions.
The physical consequences of this result are the same as the classical circulation theorem: vortices can originate inside the bulk exclusively in configurations with zero net vorticity (\eg~contractible loops in 3D or vortex-antivortex pairs in 2D).

\section{Estimates of the average vortex speed}
\label{app:vort_vel}

In Appendix~\ref{subsec:vCouples}, we presented simulations of vortex pairs in an homogeneous ($V=0$) superfluid, considering both aligned ($N_v=2$) and anti-aligned ($N_v=0$) configurations, and computed their speed $\langle v_\text{vort} \rangle$.
Here we describe how $\langle v_\text{vort} \rangle$ and its uncertainties are estimated via an average procedure over the vortex trajectories.

\begin{figure}[t]
	\centering 
	\vspace{20pt}
	\hspace{-10pt}\includegraphics[width=0.45\textwidth]{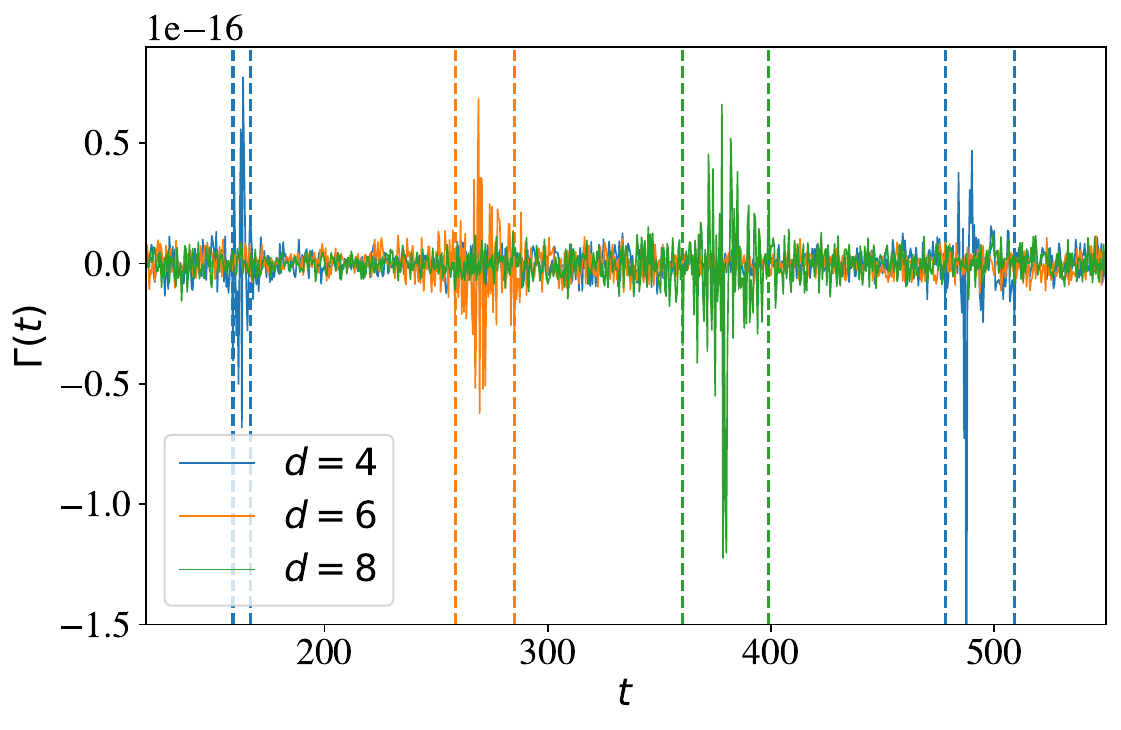}
	\caption{Circulation integrated along the boundaries of the fundamental cell for vortex–antivortex simulations with separations $d/\xi = 4, 6, 8$. The vertical dashed lines mark the time intervals where the circulation exceeds $1.8$, $2.4$, and $2.5$ times its standard deviation, chosen to identify cell crossings and avoid spurious bursts.}
	\label{fig:circ_Nv0}
\end{figure}

For $N_v=0$, we monitor the total circulation $\Gamma= \int \boldsymbol{v}\cdot d\boldsymbol{l}$ along the cell boundaries (Fig.~\ref{fig:circ_Nv0} for $d/\xi=4,6,8$). As expected, $\Gamma$ vanishes since the net vorticity is zero. Small oscillations ($\delta \Gamma \lesssim 10^{-15}$) do not affect stability, but tiny noise bursts occurring when vortices cross the boundaries can be used to estimate the average speed as $\langle v_\text{vort} \rangle \simeq (n_b+1/2)L/t_b$, with $t_b$ time of the $n_b$-th burst.
The duration of these bursts set the uncertainty on the crossing time, which is then directly propagated to $\langle v_\text{vort} \rangle$.

For $N_v=2$, we track the vortex positions in the phase and density fields. The speed is then $\langle v_\text{vort} \rangle \simeq n_r \pi d/t_r$, where $t_r$ is the time after $n_r$ full rotations of the pair. We take $\Delta t_r=10~\tau$ (four times the output resolution) as the uncertainty on $t_r$. The error bars in Fig.~\ref{fig:vCoupleVel} are propagated from these estimates.

In Appendix~\ref{subsec:gaussPeak}, we studied a vortex interacting with a Gaussian pinning potential in QPBC. There, the superfluid momentum oscillates with the same period as the vortex orbits, allowing us to extract the speed $\langle v_\text{vort} \rangle$ of the vortex around its quasi-circular orbit directly from the oscillation period.
We average the results over multiple orbits, and we take the standard deviation of the period as an estimate of its uncertainty.

\end{document}